# Is the Pale Blue Dot unique? Optimized photometric bands for identifying Earth-like exoplanets

Joshua Krissansen-Totton[1,2,3], Edward W. Schwieterman[2,3,4], Benjamin Charnay[2,3,4], Giada Arney[2,3,4], Tyler D. Robinson[2,5], Victoria Meadows[2,3,4], David C. Catling[1,2,3]

[1] Department of Earth and Space Sciences, University of Washington, Seattle, WA 98115, USA
[2] NASA Astrobiology Institute Virtual Planetary Laboratory (VPL), Seattle, WA 98115, USA
[3] Astrobiology Program, University of Washington, Seattle, USA
[4] Astronomy Department, University of Washington, Seattle, 98115, USA
[5] NASA Ames Research Center, Moffet Field, CA 94035, USA

Contact: joshkt@uw.edu

Keywords: astrobiology – planets and satellites: atmospheres – planet and satellites: gaseous planets – planets and satellites: surfaces – planets and surfaces: terrestrial planets – techniques: photometric

Short title: Is the Pale Blue Dot unique?

## ABSTRACT

The next generation of ground and space-based telescopes will image habitable planets around nearby stars. A growing literature describes how to characterize such planets with spectroscopy, but less consideration has been given to the usefulness of planet colors. Here, we investigate whether potentially Earth-like exoplanets could be identified using UV-visible-to-NIR wavelength broadband photometry (350-1000 nm). Specifically, we calculate optimal photometric bins for identifying an exo-Earth and distinguishing it from uninhabitable planets including both Solar System objects and model exoplanets. The color of some hypothetical exoplanets – particularly icy terrestrial worlds with thick atmospheres – is similar to Earth's because of Rayleigh scattering in the blue region of the spectrum. Nevertheless, subtle features in Earth's reflectance spectrum appear to be unique. In particular, Earth's reflectance spectrum has a 'U-shape' unlike all our hypothetical, uninhabitable planets. This shape is partly biogenic because $O_2$-rich, oxidizing air is transparent to sunlight, allowing prominent Rayleigh scattering, while ozone absorbs visible light, creating the bottom of the 'U'. Whether such uniqueness has practical utility depends on observational noise. If observations are photon limited or dominated by astrophysical sources (zodiacal light or imperfect starlight suppression), then the use of broadband visible wavelength photometry to identify Earth twins has little practical advantage over obtaining detailed spectra. However, if observations are dominated by dark current then optimized photometry could greatly assist preliminary characterization. We also calculate the optimal photometric bins for identifying extrasolar Archean Earths, and find that the Archean Earth is more difficult to unambiguously identify than a modern Earth twin.

## 1. INTRODUCTION

*Voyager 1's* iconic image of the pale blue dot vividly illustrates the isolation and fragility of the Earth (Sagan 1994). However, the image also shows that hints of Earth's uniqueness are visible at great distances. In both *Voyager's* image, and in more recent reincarnations such as the *Cassini* portrait (NASA 2013), the pale blue color of Earth sets it apart from any other planet in our Solar System. In this study we explore the extent to which Earth's color is unique, and evaluate whether color photometry can be used to identify Earth-like planets around other stars.

In recent years, exoplanets have been directly imaged using ground-based telescopes (Delorme et al. 2013; Kuzuhara et al. 2013; Macintosh et al. 2015; Marois et al. 2008; Rameau et al. 2013), and in some cases characterized using photometry (Janson et al. 2013; Kuzuhara, et al. 2013). Current instrumentation is limited to imaging young Jovian planets with large planet-star separations. These are still glowing with their heat of formation, which improves the planet/star contrast ratio (Macintosh et al. 2014).

However, starlight suppression technologies such as an internal coronagraph (e.g. Guyon et al. 2005; Stapelfeldt et al. 2015) or an external occulter (e.g. Cash 2006; Seager et al. 2015) could enable large-aperture future telescopes to directly image terrestrial exoplanets. In the near-term, the 2.4m WFIRST-AFTA telescope, identified as the top priority for NASA Astrophysics in the last Astronomy Decadal Survey (Blandford et al. 2010) and slated for launch in 2024, may be capable of performing the first space-based direct imaging of super-Earth planets around a small number of the very nearest stars (Spergel et al. 2015). Using the telescope/instrument noise model of Robinson et al. (2015), a 2m-class coronagraphic telescope like WFIRST-AFTA could image an Earth around a solar twin to a SNR=5 in V-band in 17 hr at 3 pc, and 76 hr at 5 pc, although this would require observing at roughly an order of magnitude below the $10^{-9}$ raw contrast, and so post-processing would be required.

The next generation large-aperture space-based telescopes concepts currently under consideration for the 2020 Decadal Survey, such as HabEx, a 4-6m-class dedicated exoplanet telescope; and the proposed, more ambitious 8-12m-class general purpose Advanced Technology Large-Aperture Space Telescope (ATLAST) (Postman et al. 2009), and High Definition Space Telescope (HDST) (Dalcanton et al. 2015), if selected for development, would be designed to be even more capable of direct imaging and spectral characterization of terrestrial exoplanets in the habitable zone. The Robinson, et al. (2015) model for a 4m-class coronagraphic telescope, with a raw contrast of $10^{-10}$ and assuming an inner working angle of $2\lambda/D$, gives 2.8 hr and 8.3 hr integration times to reach SNR=5 in V-band for an Earth around Solar twin at 3pc and 5pc respectively. For a target system at a distance of 10 pc, a 4m-class telescope can make a V-band detection in 42 hr, while a 12-m class telescope would only require 4 hr. These initial designs and calculations suggest that obtaining photometry of Earth-like planets will be readily achievable with the planned next generation telescopes.

Similarly, on the ground, the next generation 40 m-class European Extremely Large Telescope (E-ELT) has identified taking images of Earth-like exoplanets as its highest scientific priority for the telescope, and the instrumentation plan includes an ambitious and powerful planetary camera and spectrograph for this purpose (https://www.eso.org/sci/facilities/eelt/instrumentation/, see also Kasper et al. (2010)). The Thirty Meter Telescope (TMT) may also be capable of directly imaging planets in the habitable zone around M-dwarfs (Matsuo & Tamura 2010).

Once potentially habitable exoplanets are directly imaged they may then be characterized by spectroscopy. A growing literature details how the atmospheric composition and surface properties of habitable exoplanets could be observationally constrained (e.g. Cockell et al. 2009; Des Marais et al. 2002; Kaltenegger et al. 2010; Livengood et al. 2011; Robinson et al. 2010; Tinetti et al. 2006), including the possible detection of atmospheric and surface biosignatures (e.g. Kiang et al. 2007; Robinson et al. 2014; Sagan et al. 1993; Seager et al. 2012; Seager et al. 2005; Segura et al. 2003). However, for the first generation of telescopes capable of directly imaging terrestrial exoplanets, observing time devoted to exoplanets will likely be limited. The integration time required to detect an exo-Earth is significantly shorter than the time required to characterize an exo-Earth with a high-resolution spectrum. Consequently, only a handful of terrestrial exoplanets will initially be fully characterized. It would be advantageous if an initial photometric survey could identify exoplanets worthy of further characterization so telescope time can be used more judiciously. Of particular interest is whether habitable, terrestrial exoplanets can be identified using photometry alone, or whether there are photometric false positives for habitable exoplanets.

Color has been invaluable in the study of stars using the Hertzsprung-Russell diagram, but despite this precedent there is a sparse literature discussing terrestrial exoplanet characterization using photometry. Traub (2003a, 2003b) argued that planetary color might be sufficient for complete characterization and classification. Traub (2003b) used observational and modeled spectra of the Solar System planets, the Moon, and Titan to produce red and blue reflectivities on a color-color plot. It was argued that Earth occupies a unique position in red-blue color-color space. Crow et al. (2011) performed a more comprehensive analysis of the colors of Solar System bodies using photometric observations from EPOXI and other spacecraft, and also concluded that Earth occupies a unique space in red-blue color-color space compared to other Solar System bodies. However, Crow, et al. (2011) did not consider exoplanets and used fixed red and blue bands, unlike the optimization we explore in this work.

Why is Earth a pale blue dot? Rayleigh scattering is strongest at short wavelengths and in transparent atmospheres without visible-light absorbers. Because Earth has a thick and largely transparent $N_2$-dominated atmosphere it has strong reflectance in the blue (<450 nm) due to Rayleigh scattering (Crow, et al. 2011). The ocean contributes a small amount to Earth's blueness, but it is predominately a Rayleigh scattering effect (Cowan et al. 2011; Crow, et al. 2011). The Earth is pale blue rather than deep blue because white clouds enhance reflectivity at all visible wavelengths and thus reduce the relative intensity of the Rayleigh tail. Earth's reflectance spectrum is shown in Fig. 1a. Additionally, Earth's spectrum is upward sloping between 600 nm and 900 nm due to the reflectivity of continents, which includes some vegetation (red edge) component (Arnold et al. 2002; Seager, et al. 2005; Tinetti, et al. 2006). Earth's spectrum also has a slight U-shape due to the combined effect of Rayleigh scattering, upward sloping continental reflectivity, and ozone absorption in the Chappuis bands around 600 nm (Hamdani et al. 2006). The effect of the $O_3$ Chappuis bands can be seen on a daily basis. At twilight, a clear sky at the horizon would be more yellowish were it not for $O_3$ absorbing orange and yellow light, thereby producing a deep blue sky (Hulburt 1953). Chappuis bands ozone absorption is also a prominent features in Earth's transmission spectrum (Yan et al. 2015).

Crow, et al. (2011) demonstrated that Solar System planets and moons without thick atmospheres are easily distinguished from Earth because their reflectance spectra are dominated by surface absorption and scattering. Both the Moon and Mars are strongly reflective in the red due to space weathering (Lucey et al. 1995) and the presence of iron oxides (Crow, et al. 2011), respectively (see Fig. 1b). Uranus and Neptune are blue due to Rayleigh scattering, albeit to a lesser extent than Earth as a result of upper atmospheric hazes (see spectra in Fig. 1d). However, Uranus and Neptune are distinguishable from Earth on a color-color plot since they are much less reflective in the red due to methane absorption. Blueness from Rayleigh scattering is suppressed in the other Solar System bodies with thick atmospheres due to the presence of optical absorbers. This can be seen in the spectra of Jupiter, Saturn, Titan, and Venus in Fig. 1b and 1c.

Earth possesses this unique color in part due to biology. Photodissociation reactions in Earth's $O_2$-rich atmosphere generate ozone ($O_3$), and $O_3$ absorption at <340 nm generates excited oxygen, $O(^1D)$ (e.g. review by Catling 2015). In the presence of water, excited oxygen readily reacts to produce hydroxyl radicals via the following reaction (Levy 1971):

$$H_2O + O(^1D) \to 2OH \qquad (1.1)$$

These hydroxyl radicals are highly reactive and oxidize sulfur species such as $H_2S$, OCS, and dimethyl sulfate to form sulfate aerosols, which rain out of the atmosphere. Additionally, hydroxyl radicals oxidize reduced species such as CO, $CH_4$, and other hydrocarbons to form $CO_2$ and $H_2O$ (Catling 2015). The net result is that haze-forming species are destroyed, thereby maintaining a clear atmosphere.

In the absence of biogenic $O_2$, hydroxyl abundance would be low, and so an optically absorbing haze could accumulate, particularly at high altitudes above the air where rainfall operates. Furthermore, the $O_3$ absorbs in the Chappuis bands, as mentioned above, which, in combination with Rayleigh scattering and continental reflectivity, gives Earth a unique U-shaped spectrum from UV through visible wavelengths. Note that this requires both oxygen-producing life and an active hydrological cycle to produce the necessary hydroxyl radicals.

The diversity of planetary environments within the Solar System is limited, and so to determine whether habitable Earth-like planets occupy a unique position in color space we must consider possible exoplanet spectra. Due to the limitations of current instrumentation, reflectance spectra can only be obtained for transiting hot Jupiters in secondary eclipse, as has been done for HD209458b (Rowe et al. 2006) and HD189733b (Evans et al. 2013). Even in these cases, the error bars on reflectivity measurements are very large, and so we do not consider these observations in this study. For example, in the case of HD18933b, the mean reflectivity from 450-570 nm is within error of zero, and so the uncertainty in the reflectivity ratio with 450-570 nm in the denominator is effectively infinite.

In the absence of well-constrained observations, we must rely upon atmospheric models to determine the likely colors of exoplanets. There have been numerous attempts to generate plausible reflectance spectra of exoplanets. Sudarsky et al. (2000) modeled the reflectance spectra of gas giants by specifying composition and applying a radiative-transfer model that incorporated condensables. Five distinct giant planet classes were identified on the basis of temperature. Spiegel et al. (2010) modeled the atmospheres of Neptune-sized exoplanets using a similar methodology, and also demonstrated that planet-star separation (temperature) has a strong effect on color because the presence of both highly reflective condensables and short-wave absorbers is temperature dependent. Cahoy et al. (2010) modeled extrasolar Jupiter and Neptune analogs for a range of star-planet separations, viewing phases, and metallicities. They concluded that viewing phase may have strong effect on Jovian reflectance spectra, which can make photometric characterization challenging.

The colors of terrestrial exoplanets have also been modeled with special attention given to the possible surface spectra of terrestrial exoplanets. For instance Sanromá et al. (2013) modeled the reflectance spectra of Earth-like planets with different continental surfaces such as desert, microbial mats, and vegetation, and Sanromá et al. (2014) calculated Archean Earth spectra for coastal areas dominated by purple bacteria. Hegde and Kaltenegger (2013) and Hegde et al. (2015) also described reflectance spectra for different surface features including extremophiles. However, the neglect of an atmosphere in Hegde and Kaltenegger (2013) means that the reflectance spectra and their derived broadband colors are not what would be observed remotely. Schwieterman et al. (2015a) measured the reflectance spectra of a diverse range of microorganism colonies and found that a halophile world might be remotely identifiable. Meadows (2006) and Kaltenegger et al. (2007) modeled the evolution of Earth's reflectance spectrum from the Archean to the present as an analog for exoplanet observations. Fig. 1e shows a selection of model exoplanet spectra.

Despite this abundance of model spectra, there has not been a systematic study of exoplanet reflectance spectra to determine the extent to which terrestrial planets can be distinguished from giant planets on the basis of color, or the extent to which Earth-like "pale blue dots" occupy a unique location in color-color space taking into account possible false positives. In this study we use both model and observed reflectance spectra to explore the locations of planets in color-color space. Specifically, we calculate the optimal photometric bands for identifying Earth-like exoplanets and distinguishing them from uninhabitable planets. Our sample of uninhabitable planets is not intended to be exhaustive, but the model and observed spectra we consider are broadly representative of the types of planets that could be mistaken for habitable Earth-like planets, based on current knowledge. We also consider the prospects for identifying extrasolar Archean Earths using photometry. Finally, we compare the relative telescope integration time required to identify an exo-Earth using color to the time required to obtain a spectrum.

## 2. METHODS

*2.1 Reflectance spectra*

The visible reflectance spectra used in this study were obtained from a variety of sources and are summarized in Table 1. Here, we describe in detail how model reflectance spectra were generated and where observed spectra were sourced.

For most Solar System objects, we used spectral observations from spacecraft and ground-based telescopes. Reflectance spectra for Jupiter, Saturn, Uranus, Neptune and Titan were obtained from Karkoschka (1998), and reflectance spectra for Callisto, Io, Europa and Ganymede were obtained from Karkoschka (1994). All of these spectra are full phase or very close to full phase (<7°).

The model reflectance spectra used in this study were generated using the Spectral Mapping Atmospheric Radiative Transfer Model (SMART). SMART is a one-dimensional, plane-parallel, line-by-line, multiple-scattering radiative transfer model developed by D. Crisp (Crisp 1997; Meadows & Crisp 1996). SMART takes as input vertical profiles of temperature, pressure, gas mixing ratios, and aerosol optical depths. With these inputs SMART calculates the monochromatic optical properties of each layer. Line absorption cross-sections are calculated with a program called Line-by-Line Absorption Coefficients (LBLABC), which determines the absorption coefficients of spectrally active given a line list database (Crisp 1997; Meadows & Crisp 1996). Because we leverage many existing spectra from the literature, the older HITRAN 2008 (Rothman et al. 2009) line lists were used to calculate the absorption coefficients for the spectrally absorbing gases in our model atmospheres except where stated otherwise. Throughout this work, when calculating gaseous absorption coefficients with LBLABC, a line cutoff of 1000 $cm^{-1}$ is used. The gas absorption cross-sections determined by LBLABC are read in by SMART and multiplied by the gas mixing ratios and layer thicknesses to determine the normal-incidence gas extinction optical depths. The total extinction at each layer and each hyperfine spectral grid point is calculated by combining the gas extinction optical depths with those calculated for Rayleigh scattering and aerosols. We used a single solar zenith angle of 60 degrees for all runs, which approximates the illumination observed in a planetary disk average during an observation at quadrature phase (Segura et al. 2005). We assume the surfaces in this model are Lambertian reflectors, which is an approximately true assumption for the Earth at non-crescent phase angles (Robinson, et al. 2010).

Earth spectra were obtained from Robinson *et al.* (2014; 2011). We used a validated 3D Earth model since this enabled us to explore how Earth's reflectance spectrum changes with phase. Validated (1D) model spectra were also used for Mars, Venus, and the Moon (Robinson, et al. 2011). The disk-averaged Mars surface albedo spectrum was the same as that used in Crisp (1990) and was constructed from data obtained from McCord et al. (1982) and Blaney et al. (1988)

A coupled 1D photochemical-climate model was used to generate self-consistent ancient Earth atmospheres, both with and without haze, that are best estimates of those that existed in the Archean Eon (4.0-2.5 Ga). The radiative-convective climate portion of the model is based on the code developed originally by Kasting and Ackerman (1986) and updated most recently to study habitable zone boundaries (Kopparapu et al. 2013).

The photochemical portion of the coupled code originated in Kasting et al. (1979) and has since been modified extensively, including in a study that examined the potential for hazes to exist in the atmosphere of early Earth (Zerkle et al. 2012) using the mean field approximation for fractal particle scattering (Botet et al. 1997). The photochemical model includes 200 plane-parallel layers with a layer spacing of 0.5 km up to an altitude of 100 km. The abundances of chemically active species in each layer are calculated by solving mass and flux continuity equations using a reverse-Euler method. Outputs from this coupled model (haze particle density, the temperature profile, gas mixing ratios) are input into SMART to generate spectra. Haze refractive indices are taken from Khare et al. (1984).

The Archean Earth may have had a hydrocarbon haze in its atmosphere during some intervals (Domagal-Goldman et al. 2008; Pavlov et al. 2001; Zerkle, et al. 2012). This fractal hydrocarbon haze thickness scales with increasing $CH_4$ relative to $CO_2$ (Haqq-Misra et al. 2008). Our hazy Archean simulations have 1% $CO_2$, below a recent paleosol upper limit (Driese et al. 2011) for 2.7 Ga. A $CH_4/CO_2$ ratio of 0.2 generates a substantial haze in the coupled photochemical-climate model. A thicker haze, with larger particles, may be possible at a higher $CH_4/CO_2$ ratio or a hotter atmosphere because the particle coagulation timescale depends on temperature (Tolfo 1977). For this thicker haze scenario, we use a spectrum with $CH_4/CO_2 = 1$. For the purposes of this study, the $CH_4/CO_2 = 0.2$ spectrum is considered a "thin" haze compared to the "thick" haze of large particles generated by $CH_4/CO_2 = 1$. The solar constant in these simulations is scaled to 2.7 Ga and includes a wavelength-dependent scaling that corrects for solar evolution (Claire et al. 2012). The haze-free Archean Earth atmosphere was assumed to have 50% clear skies, 25% high cloud (8.5km), and 25% low cloud (1.5 km), which is a cloud parameterization used for the modern Earth (Robinson, et al. 2011). Hazy spectra include the same cloud parameterization along with global coverage by hydrocarbons. A full description of the methodology used to generate Archean Earth atmospheres can be found in Arney et al. (2015). Note that HITRAN 2012 (Rothman et al. 2013) was used to generate Archean Earth absorption coefficients.

To investigate the diversity of colors of giant planets, we included a number of model spectra of giant exoplanets and sub-Neptunes. Giant planet reflectance spectra were obtained from Sudarsky, et al. (2000). Sudarsky, et al. (2000) and Sudarsky et al. (2005) define five classes of giant planets based on temperature. Class 1 giant planets are cool (<150 K) and highly reflective due to ammonia clouds, class 2 planets are warmer (250 K) and reflective due to water clouds, class 3 and 4 planets are warmer still (≥350K and 900-1500 K respectively) and strongly absorbing due to alkali absorption bands and the absence of condensable species in the upper atmosphere, and class 5 planets are the hottest (>1500 K) and highly reflective due to silica and iron clouds. The bulk compositions of all five classes are assumed to be the same; what determines any giant planet's class is its planet-star separation (i.e. temperature). The effects of metallicity on reflectance spectra are relatively minor compared to the effects of temperature (Cahoy, et al. 2010). We only consider class 1 and 2 giant planets in this study since higher temperature gas giants would necessarily be well interior to the inner edge of the habitable zone and thus easily distinguishable from habitable worlds on the basis of orbital distance.

NASA's *Kepler* mission has revealed that Super-Earths and sub-Neptunes are common (Fressin et al. 2013; Howard et al. 2010; Marcy et al. 2014). To better sample the diversity of exoplanet colors we generated a number of plausible sub-Neptune reflectance spectra based on a planet with the same mass (6.6 $m_\oplus$) and radius (2.4 $R_\oplus$) as GJ1214b, but with a stellar insolation of 2000 W/m². Reflectance spectra were generated for a cloud free $H_2$ dominated atmosphere with solar elemental abundances and a $H_2$ dominated atmosphere with 100x solar metallicity (solar nebula atomic abundances from Lodders (2003)). Each atmosphere was assumed to be in chemical equilibrium, and the temperature pressure profiles were obtained using the 1D version of the Generic LMDZ GCM (Charnay et al. 2015; Wordsworth et al. 2011). We also computed the reflectance spectra for a sub-Neptune with a pure steam atmosphere (water clouds were neglected). Note that HITRAN 2012 (Rothman, et al. 2013) was used to generate sub-Neptune absorption coefficients.

The model spectra for the giant planets and sub-Neptunes described above are unrealistic because they assume that each atmosphere is in chemical equilibrium and so they do not include the effects of photochemical hazes. In our Solar System, both Jupiter and Saturn possess photochemical hazes that affect their color by suppressing Rayleigh scattering (Crow, et al. 2011). Sudarsky, et al. (2000) and Spiegel, et al. (2010) considered the effects of adding photochemical hazes to their giant planet models, and we used the hazy class 1 giant planet reflectance spectra from Sudarsky, et al. (2000) in this study. The model spectra for giant planets and sub-Neptunes used in this study are speculative because currently few observational constraints exist for the reflectance spectra of exoplanets.

To simulate the color of a snowball Earth we used reflectance spectra for snow and blue ice surfaces from Warren et al. (2002) and Grenfell et al. (1994), as cited in Shields et al. (2013). We assumed an Earth-like atmosphere above these surfaces, and computed the resultant reflectance spectra for both a 100% snow and 100% blue ice planet. We also computed the reflectance spectra for an Earth-like atmosphere above a 100% kaolinite surface. These three spectra can be considered as end-member cases for a dusty snowball Earth. In our color-color plots we assume that a triangle region bounded by these three spectra encompasses the possible colors of a snowball Earth (see results section).

*2.2 Optimization algorithm*

For the purposes of this paper, we define a habitable planet as a terrestrial planet within its star's habitable zone (Kasting et al. 1993; Kopparapu, et al. 2013). Firstly, we consider the color of Earth and all uninhabitable planets, where the latter set includes gas and ice giants, hydrogen dominated and thick steam atmosphere sub-Neptunes, permanent snowballs, and all Solar System bodies excluding Earth. Table 1 indicates which planets are assumed to be uninhabitable. We refer to this set of uninhabitable planets as potential false positives, and we seek the photometric bins that optimally separate Earth from this set of false positives in color-color space such that probable exo-Earths can be easily identified by photometric observations.

The reflectivity of a planet, $R_{[a,b]}$, is defined as the mean reflectivity in the wavelength range $a$ to $b$ (in nm). In the simplest case, we seek three photometric bins $\{[a_1,b_1],[a_2,b_2],[a_3,b_3]\}$ which maximally separate Earth from false positives in color-color space, where $[a_i,b_i]$ denotes the wavelength range (in nm) of the $i$th photometric bin. Given these three bins, the reflectance spectra of every planet is calculated for each bin, and the planet can be plotted in color-color space, where color axes are defined as the reflectivity ratios $R_{[a_3,b_3]}/R_{[a_2,b_2]} = R_{red}/R_{green}$ and $R_{[a_1,b_1]}/R_{[a_2,b_2]} = R_{blue}/R_{green}$. Note that whenever reporting optimizations involving three bins, we typically use "red" to label the color of the longest wavelength bin, "blue" for the shortest wavelength bin, and "green" for the intermediate bin. These "red", "green", and "blue" labels may not correspond to the true colors of the bin. We use reflectivity ratios rather than absolute reflectivity because radius-albedo degeneracy will typically preclude the accurate measurement of absolute reflectivity from direct imaging alone.

We applied the optimization algorithm Constrained Optimization by Linear Approximation (COBYA) in Python to determine the set of photometric bins that maximizes Earth's separation from false positive planets, subject to constraints ensuring non-overlapping bins that fit within the visible spectrum and are at least 100 nm wide (unless stated otherwise). The quantity that was maximized was the Pythagorean distance between Earth and the nearest false positive neighbor in color-color space. The optimization problem can be formally posed as follows:

$$\underset{a_1,a_2,a_3,b_1,b_2,b_3}{maximize} \quad f(a_1,a_2,a_3,b_1,b_2,b_3) =$$

$$\min\left(\sqrt{\left[\left(R_{[a_1,b_1]}/R_{[a_2,b_2]}\right)_{Earth} - \left(R_{[a_1,b_1]}/R_{[a_2,b_2]}\right)_j\right]^2 + \left[\left(R_{[a_3,b_3]}/R_{[a_2,b_2]}\right)_{Earth} - \left(R_{[a_3,b_3]}/R_{[a_2,b_2]}\right)_j\right]^2}\right) \quad (1.2)$$

$$subject\ to \quad \{\lambda_{min} < a_1\}, \{b_3 < \lambda_{max}\}, \{a_i + m < b_i\}, \{b_i < a_{i+1}\}, \quad i = 1, 2, 3$$

Here, $j$ is a member of the set of false-positive (uninhabitable) planets (see Table 1). The smallest wavelength used to form photometric bins, $\lambda_{min}$, is typically 350 nm in our analysis. Similarly, $\lambda_{max} = 1000$ nm is the maximum allowable wavelength used to construct photometric bins. The minimal photometric bin width, $m$, is typically taken to be 100 nm unless stated otherwise. The first and second inequality constraints ensure that all three photometric bins fit into the desired spectral range (350-1000 nm). The third inequality constraint enforces the minimum bin size, and the fourth inequality constraint ensures that bins do not overlap.

The function to be maximized in (1.2) is not convex, and so local maxima are not guaranteed to be the global maximum. Consequently, we implemented a brute force global optimization search by iterating the COBYA algorithm over 2000 randomized initial conditions and selecting the best solution as the global maximum. Repeat calculations consistently produced the same answer, thereby establishing that this optimization algorithm with 2000 iterations successfully computes the global maximum.

This methodology assumes three photometric bins with "red"/"green" and "blue"/"green" axes, but extension to three or four bins with generalized axes was also done for comparison. Rather than force the horizontal axis to be "red"/"green" and the vertical axis to be "blue"/"green" reflectivity ratios, the optimization algorithm can be modified to allow for any combination of bins on each axis (e.g. $R_{[a_i,b_i]}/R_{[a_k,b_k]}$ and $R_{[a_l,b_l]}/R_{[a_n,b_n]}$). In other words, calculations with generalized axes do not force the numerator (denominator) on the vertical axis to be the shortest (middle) wavelength bin, and do not force the numerator (denominator) on the horizontal axis to be the longest (middle) wavelength bin.

### 3. RESULTS

*3.1 Identifying exo-Earths*

The color-color plot with optimized bins for separating Earth from false positives is shown in Fig. 2. The blue circle shows the Earth's position in color-color space (at quadrature), and the remaining symbols are the set of uninhabitable false positives. Some false positives are directly labeled and some are labeled in the legend, depending on available space. Earth is the only planet in the set we considered where both the "blue"/"green" and "red"/"green" ratios are appreciably bigger than 1. The wavelength ranges for the three optimal bins that define the reflectivity axes are 431-531 nm, 569-693 nm, 770-894 nm, and Earth is separated from its nearest neighbor (Callisto with a 0.75 bar $N_2$ atmosphere) by a distance of 0.32 in color-color space. The implications of the dimensionless separation are discussed below. A complete list of all the false positives in Fig. 2 is available in Table 1.

The red shaded region in Fig. 2 denotes the possible range of colors for a snowball Earth planet. This region encloses linear combinations of blue ice, snow, and kaolinite (dust) to represent spectra for a dirty snowball, as described in the methods section. The snowball false positive is assumed to possess an Earth-like atmosphere with clear skies. If clouds are added to the snowball then it moves slightly closer to Earth in color-color space. No false positives from our sample of planets occupy the exact same region of color-color space as the Earth (Fig. 2).

Many of the false positives we considered are models of icy worlds with relatively thick atmospheres. For instance, we computed the reflectance spectra for hypothetical exoplanets with surface spectra from Callisto, Europa, Ganymede, and Io, overlain by thick nitrogen atmospheres with surface pressures ranging 0.5-2 bar. Additionally, we computed the reflectance spectra for a hypothetical exoplanet with a Mars-like surface spectrum and a thick atmosphere of either $N_2$ or $CO_2$. Finally, we modeled thick atmospheres overlying kaolinite, $CO_2$ ice, and a gray surface to represent the possible diversity of icy worlds. Unless stated otherwise we assumed 250 K isothermal atmosphere for all icy planet models because this temperature is consistent with a frozen, uninhabitable surface, but is not so different from habitable conditions that it would be easily distinguishable on the basis of semi-major axis. Note however that the temperature does not affect the spectrum, only the implied atmospheric mass. These planets represent uninhabitable false positives that could plausibly be found beyond the outer edge of the habitable zone (semi-major axis may initially be unknown for planets discovered by direct imaging surveys). They were included in this study because the strong Rayleigh scattering from their thick, clear atmospheres, in combination with relatively flat or upward sloping surface spectra makes them similar in color to Earth. Some of these false positive spectra are shown in Fig. 3b and 3c.

Earth is separated from all the false positives due to the unique U-shape of its spectrum. Fig. 3a shows the Earth's reflectance spectrum superimposed on the optimal photometric bins that define the axes in Fig. 2. Here, it can be seen that the optimization algorithm picked bins that capture the U-shape of the Earth's spectrum. As described in the introduction, the high reflectivity in the blue is largely attributable to Rayleigh scattering, whereas the upward sloping shape of the spectrum between 600 and 900 nm is attributable to the reflectivity of continents, which includes some vegetation (red edge) component. (Tinetti, et al. 2006). Reduced reflectivity around 600 nm due to ozone absorption in the Chappuis bands also contributes to the U-shape. This can be seen in Fig. 3a where we plot Earth's spectrum without ozone alongside Earth's spectrum with ozone. Ozone absorption is responsible for the bottom of the U-shape, and without ozone Earth would occupy a somewhat different position in color-color space.

If the optimization calculation is repeated - but the icy false positives are excluded - then the separation between Earth and its closest (non icy false positive) neighbor increases. This is shown in Fig. 4 where the separation between Earth and its nearest neighbor (a Class 1 giant, i.e. cool with ammonia clouds) in color-color space is 0.53. Here, the wavelength ranges for the three optimal bins that define the reflectivity axes are 350-450 nm, 593-693 nm, and 879-992 nm. This set of photometric bins does a better job of separating Earth from the non-icy bodies than the photometric bins in Fig. 2. However, if these revised photometric bins were used to identify potential exo-Earths then any exo-Earth would be indistinguishable from uninhabitable icy worlds. Fig. 4 shows that with these photometric bins Earth occupies the same region in color-color space as a planet with a 2 bar $CO_2$ atmosphere overlying Mars' surface or a planet with a 0.75 bar $N_2$ atmosphere overlying Callisto's surface. Even a planet with a gray (flat) surface spectrum and a 1-2 bar clear $N_2$ atmosphere is very similar in color to Earth. In short there are many ways to make a quasi pale blue dot that is not habitable.

For the purposes of characterizing exo-Earths, it is beneficial to allow gaps between photometric bins. For example, if the optimization algorithm is modified to disallow gaps between bins by enforcing the additional constraint, $\{b_i = a_{i+1}\}$, then the optimal separation between Earth and its nearest neighbor is reduced to 0.27 (not shown).

In addition to considering the forced "red"/"green" and "blue"/"green" axes in Fig. 2 and 4, we also optimized the separation between Earth and false-positive neighbors in color-color space with generalized axes. In other words, we allowed the optimization algorithm to consider all possible combinations of reflectivity ratios for use on the horizontal axis and vertical axis of color-color space. Fig. 5 shows the optimal separation between Earth and its nearest neighbor using generalized axes and three photometric bins. In this case the three bins that define the axes are 579-679 nm ("yellow"), 797-897 nm ("red"), and 898-999 nm ("far red"), and the color-color space axes for this optimal separation are $R_{[797,897]} / R_{[579,679]}$ and $R_{[797,897]} / R_{[898,999]}$. Here, Earth is separated from its nearest neighbor (Mars) by 0.40 in color-color space; this is a marginal improvement over the separation using "red"/"green" and "blue"/"green" axes. The vast majority of planets have a $R_{[797,897]} / R_{[898,999]}$ ratio of approximately 1, but Earth is closer to 1.5 due to strong water absorption around 940 nm. Thus, although generalized axes do a slightly better job of separating the Earth from uninhabitable planets than the forced "red"/"green" and "blue"/"green" axes, Earth is separated due to the presence of water (a habitability feature) and not specifically biogenic features of its spectrum. If the calculation is repeated using four photometric bins rather than three, then the optimal separation is 0.38 (not shown). This demonstrates there are no added benefits to using four photometric bins rather than three photometric bins to identify exo-Earths.

We also investigated optimizing Earth's separation from false-positive planets in a 3-dimensional color space, and by modifying the separation criteria to be optimized. Examples of the latter included maximizing the mean or median separation between Earth and all its neighbors, maximizing the separation to the nearest neighbor scaled by the mean difference between all pairs of planets, and fitting a linear regression to all planets and maximizing Earth's residual. None of these approaches were improvements over the simple criteria applied in figures 2, 4 and 5.

*3.2 The effect of phase*

All the model spectra so far in this study represent observations during quadrature, which is the most probable viewing geometry for directly imaged exoplanets. Fig. 6 shows the probability distribution of apparent phase for the direct imaging of an exoplanet with a random inclination and position in its orbit. This was calculated by repeatedly sampling uniform distributions of inclination and position in orbit, and calculating the resultant phase as viewed by a distant observer. Disk-averaged quadrature observations do not exist for many Solar System objects, and so we were forced to use full phase observations in these cases. It is beyond the scope of this study to model the 3D phase-dependent photometry for every planet we considered. However, it is possible to explore the effects of phase for some planets to see how this affects our conclusions.

A validated 3D Earth model was used to generate Earth's reflectance spectrum (see methods), and so the effect of phase on Earth's position in color-color space can be explored. Fig. 7 is identical to Fig. 2 (Earth optimally separated from all false positives), except that we have plotted four blue symbols for the Earth representing full, quadrature, gibbous, and crescent phases. Evidently, viewing phase does not have a large effect on Earth's color – all four phases are clearly separated from the uninhabitable false positives in color-color space. This is a general result that is not dependent on the specific photometric bins used.

Cahoy, et al. (2010) used an albedo spectral model with detailed scattering calculations to compute the reflectance spectra of Jupiter and Neptune analogs as a function of orbital phase. In this model the emergent flux is integrated over the planet's surface to capture spatially-dependent scattering behavior. It is found that the Jupiter analog – a 3x solar metallicity Jovian at 0.8 AU with an atmosphere in chemical equilibrium – has the most phase variability in its reflectance spectrum. The Jovian model spectra from figure 12 of Cahoy, et al. (2010) are plotted in color-color space in our Fig. 7 to show an extreme example of phase-dependent color. Here, the "blue"/"green" ratio decreases dramatically as phase angle increases. This is analogous to the sunset effect: at large phase angles (crescent phase), the mean path length through the atmosphere is large, and so blue light is preferentially scattered out of the observer's beam.

### 3.3 The effect of clouds

The clouds for the Earth spectra used in our analyses are parametrized based on the comparison of *EPOXI* Earth observations to the 3D Earth model (Robinson, et al. 2011). Exo-Earths could have different levels of cloudiness to the modern Earth, and so the effects of cloudiness on color need to be evaluated. We investigated the dependence of Earth's color on cloud fraction and height by computing a 1D Earth spectrum for five different cases: 100% low level stratocumulus cloud (1.5 km, optical depth 10), 100% medium-level alto-stratus clouds (3.9 km, optical depth 10), 100% high level cirrus clouds (8.5 km, optical depth 10), clear skies, and clear skies with a 100% ocean surface. These end-member cases are plotted in color-color space in Fig. 7. The blue shaded region spans the range of colors possible for different levels of cloudiness. We find that 100% cloud cover at any height moves the Earth closer to the icy false positives in color-color space. This is unsurprising because 100% cloud cover eliminates all contributions from the surface reflectance spectrum, which is responsible for the upward sloping shape in the red portion of Earth's spectrum. Additionally, 100% cloud cover reduces the average path length in the atmosphere, which diminishes the blueness due to Rayleigh scattering. This explains why the 'clear skies with a 100% ocean surface' end-member is bluer than all the other Earth spectra, although the ocean surface itself also contributes to the blueness (Crow, et al. 2011). The position of the clear sky Earth shows the redness of the vegetated continental surface compared to the ocean.

### 3.4 Minimum bin size

There is a trade-off between optimal color-color separation and photometric bin sizes. Reducing the minimal bin size ($m$) will result in larger optimal separations between Earth and its nearest neighbor because narrower, more diagnostic regions of the spectra can be targeted. However, reducing the minimal bin size will reduce the photon flux in each bin, thereby increasing the integration time required to separate the Earth from its nearest neighbor to any desired level of precision. To investigate the effect of minimum bin size, we repeated the Earth-separating optimization and varied the minimum bin size from 50 nm to 150 nm. This is shown in Fig. 8 where the optimal separation between Earth and its nearest neighbor is plotted as a function of minimum bin size. Reducing the minimum bin size does not improve the optimal separation much. For a minimum bin size of 50 nm, the optimal separation between Earth and its nearest neighbor is 0.39. This is only 20% larger than the optimal separation for 100 nm minimum bins (0.32). The minimum bin size must be decreased to 30 nm before the nearest neighbor separation increases markedly to 1.10 (not shown).

### 3.5 Identifying extrasolar Archean Earths

If the Archean Earth is plotted in color-color space using the three photometric bins that optimally separate the modern Earth, then it is clear that the Archean Earth is difficult to identify (Fig. 7). For example, the hazy Archean Earth could be mistaken for a gas giant or Titan-like exoplanet, whereas the non-hazy Archean Earth could be mistaken for a Mars-like exoplanet with a thick atmosphere. This demonstrates the potential for false negatives: planets that appear to be uninhabitable based on photometry, but could in fact be revealed as habitable using high resolution spectroscopy.

To try to distinguish Archean Earths, we repeated the optimal separation calculations described above with the Archean Earth as the subject. Specifically, we repeated the calculations four times to represent four plausible Archean Earth scenarios: thick-haze with high methane (1% mixing ratio), thin-haze with low methane (0.2% mixing ratio), non-hazy with high methane (1% mixing ratio), and non-hazy with low methane (0.2% mixing ratio). Note that although the modern Earth is included in the figures that follow, we only attempted to separate the Archean Earth from the set of false positives: the modern Earth is merely plotted as a reference.

It is extremely difficult to uniquely identify the hazy Archean Earth using photometry alone. Fig. 9a shows the thin-haze Archean Earth optimally separated from all potential false-positives using the "red"/"green" and "blue"/"green" axes. Here, the optimal photometric bins are 350-535 nm ("blue"), 780-893 nm ("green"), and 900-1000 nm ("red"), and the Archean Earth is separated from its nearest neighbor, Jupiter, by only 0.20 in color-color space (note that the scaling of the horizontal axis is bigger than the vertical axis). Modifying the optimization algorithm to allow for generalized axes improves the optimal separation slightly, but tends to place the Archean Earth in a non-unique region of color-color space. For example Fig. 9b shows the thick haze Archean Earth optimally separated from all false-positives using generalized three-bin axes. In this case, the optimal photometric bins are 350-450 nm ("blue"), 572-853 nm ("green"), and 853-953 nm ("red") with corresponding axes $R_{[572,853]}/R_{[350,450]}$ and $R_{[572,853]}/R_{[853,953]}$. The separation between the Archean Earth and its nearest neighbor is 0.44, but Fig. 9b shows that it is positioned between Jupiter and Saturn; it is easy to imagine an intermediate gas giant that could be conflated with the thick-hazed Archean Earth using these photometric bins.

It is also difficult to identify the non-hazy Archean Earth using photometry. For example Fig. 9c shows the high methane non-hazy Archean Earth optimally separated from all false positives using generalized reflectivity axes with three photometric bins. The optimal photometric bins are 350-450 nm ("blue"), 504-625 nm ("green"), and 879-991 nm ("red") with corresponding axes $R_{[879,991]}/R_{[504,625]}$ and $R_{[350,450]}/R_{[879,991]}$. The separation between the Archean Earth and its nearest neighbor (100% snow) is 1.47. Although this separation is comparatively large, the extent to which the non-hazy Archean Earth occupies a unique position in color-color space is once again unclear; for example, an intermediate between Jupiter and Uranus (or the sub-Neptune model) could potentially occupy the same region in color-color space. If the calculations are repeated for 4 photometric bins, or for the low methane (2000 ppmv) non-hazy Archean Earth case then similar results are obtained (not shown). In short, the non-hazy Archean Earth can be separated from false positives, but it does not occupy a unique region in color-color space. We did not consider the oxygenated Proterozoic Earth because previous work shows that its spectrum is similar to the modern Earth (Kaltenegger, et al. 2007; Meadows 2006). However, the Chappuis bands - and therefore their contribution to Earth's unique U-shape - are not visible in both Proterozoic and Archean Earth spectra. This suggests that for most of Earth's history, it was more difficult to photometrically distinguish from uninhabitable planets than the modern Earth.

*3.6 Detectability considerations*

Using optimized photometric bins, we have shown that (modern) Earth can be separated from false positives in color-color space by around 0.3-0.4 in a particular dimensionless color-color space. Furthermore, there is reason to believe that Earth occupies a unique region in color-color space due to the U-shape of its spectrum from far UV to near-IR. Whether or not this unique color has any utility depends on how difficult it is to constrain color through photometric observations. Rather than calculate the necessary telescope integration time required to identify an exo-Earth, we calculate a dimensionless ratio of the integration time required to identify an exo-Earth using photometry relative to the integration time required to obtain a spectrum of the same exo-Earth. This ratiometric approach does not require any explicit assumptions about telescope parameters or a detailed noise model, and so the results will apply to any future telescope.

Specifically, we wish to compare the telescope integration time required to separate Earth from its nearest false positive with 5σ precision to the integration time required to obtain a spectrum with signal-to-noise of 5. Fig. 10 shows a zoomed-in version of Fig. 2, which highlights the important parameters for this calculation. We seek the integration time that ensures the separation distance, $S_d$, between exo-Earth and the nearest neighbor is five times larger the uncertainty in this separation, $\sigma_s$.

The separation distance is given by $S_d = \sqrt{x^2 + y^2}$, where $x = (R_{red}/R_{green})_{Earth} - (R_{red}/R_{green})_{Neighbor}$ is the distance between Earth and its nearest neighbor along the horizontal axis, and $y = (R_{blue}/R_{green})_{Earth} - (R_{blue}/R_{green})_{Neighbor}$ is the distance between Earth and its nearest neighbor along the vertical axis. The uncertainty in the total separation is given by:

$$\sigma_S = \sqrt{\left(\frac{\partial S_d}{\partial x}\sigma_x\right)^2 + \left(\frac{\partial S_d}{\partial y}\sigma_y\right)^2} = \sqrt{\frac{x^2\sigma_x^2 + y^2\sigma_y^2}{x^2 + y^2}} \tag{1.3}$$

where $\sigma_x$ and $\sigma_x$ are the uncertainties in $x$ and $y$, respectively.

The reflectivity measurements $R_{blue}$, $R_{green}$ and $R_{red}$ will each have an associated observational uncertainty, $\sigma_{blue}$, $\sigma_{green}$ and $\sigma_{red}$, respectively (for an exo-Earth). These can be recast in terms of the signal-to-noise $(S/N)$ reciprocal for each photometric bin ($\sigma_{blue}/R_{blue} = (N/S)_{blue}$ etc.). Assuming there is no uncertainty in the position of the nearest neighbor, the rules of error propagation then allow us to express the $S_d/\sigma_S$ ratio as follows, using equation (1.3):

$$\frac{S_d}{\sigma_S} = \frac{x^2 + y^2}{\left(x^2\sigma_x^2 + y^2\sigma_y^2\right)^{1/2}}$$

$$= \frac{\left(x^2 + y^2\right)}{\sqrt{x^2\left((R_{red}/R_{green})_{Earth}\sqrt{\left(\frac{N}{S}\right)_{red}^2 + \left(\frac{N}{S}\right)_{green}^2}\right)^2 + y^2\left((R_{blue}/R_{green})_{Earth}\sqrt{\left(\frac{N}{S}\right)_{blue}^2 + \left(\frac{N}{S}\right)_{green}^2}\right)^2}} \tag{1.4}$$

The noise-to-signal $(N/S)$ for each photometric bin will be a monotonically decreasing function of the telescope integration time, which is in turn dependent on the stellar spectrum, Earth's wavelength dependent reflectivity, external noise sources such as exozodiacal dust, and various telescope parameters. For simplicity, we will assume that $N/S$ for all three bins is the same for any given integration time i.e. $N/S = (N/S)_{red} = (N/S)_{green} = (N/S)_{blue}$. This is approximately true since neither stellar flux nor Earth's albedo varies greatly across the visible spectrum. By making this assumption and substituting in the values from Fig. 10 we have:

$$\frac{S}{\sigma_S} = \frac{\left(0.31^2 + 0.079^2\right)}{\sqrt{0.31^2\left(1.25\sqrt{2\left(\frac{N}{S}\right)^2}\right)^2 + .079^2\left(1.27\sqrt{2\left(\frac{N}{S}\right)^2}\right)^2}} = 0.181\left(\frac{S}{N}\right) \quad (1.5)$$

Since we wish to separate Earth to 5σ the signal-to-noise in each individual photometric bin must be $S/N = 5/0.181 = 27.6$. Incidentally, this implies that photometric observations must be accurate to within 1/27.6 ≈ 4% to separate Earth from uninhabitable planets to 5σ.

We consider two endmember cases for the noise source in direct imaging telescope observations. Firstly, we assume the signal-to-noise is dominated by astrophysical noise such as zodiacal light, imperfectly suppressed starlight, or is photon limited. The signal-to-noise for any such observation is given by:

$$S/N = \frac{F_{Planet}t}{\sqrt{F_{Planet}t + F_{Zodiacal}t + F_{null}t}}$$
$$= \frac{F_{Planet}}{\sqrt{F_{Planet} + F_{Zodiacal} + F_{null}}}\sqrt{t} \quad (1.6)$$

Here $F_{Planet}$ is the photon flux from the planet in the wavelength range being observed, $F_{Zodiacal}$ is the exo-zodiacal and zodiacal photon flux in the wavelength range being observed, $F_{null}$ is the unsuppressed starlight in the wavelength range being observed, and $t$ is the integration time.

The integration time required, $t_{color}$, to obtain a S/N of 27.6 in each 100 nm reflectivity bin (and therefore separate Earth from its nearest neighbor by 5σ) will be given as follows, using equation (1.6):

$$27.6 = S/N = \frac{F_{Planet}}{\sqrt{F_{Planet} + F_{Zodiacal} + F_{null}}}\sqrt{t_{color}} \quad (1.7)$$

This photometry integration time should be compared to the integration time required to obtain a $S/N = 5$ spectrum with 10 nm bins. In effect, we are comparing the time required to photometrically separate the Earth to 5σ precision to the time required to constrain the reflectivity at every point in a spectrum with the same level of precision. We chose 10 nm bins to achieve a moderate spectral resolution of 70 at 700 nm (λ/Δλ=700/10=70). The integration time required to obtain such a spectrum, $t_{spectrum}$, is given by:

$$5 = S/N = \frac{F_{Planet}'}{\sqrt{F_{Planet}' + F_{Zodiacal}' + F_{null}'}}\sqrt{t_{spectrum}} \quad (1.8)$$

Since the spectral bins (10 nm wide) are 10 times smaller than the photometric bins (>100 nm wide), then it will be approximately true that all fluxes decrease by a factor of 10. In other words $10F_X' = F_X$ for all the fluxes in equation (1.8). By combining equations (1.7) and (1.8) we arrive at the result:

$$t_{spectrum} \approx 0.33 t_{color} \quad (1.9)$$

This demonstrates that *if observational noise is dominated by astrophysical sources then the integration time required to separate Earth from the nearest false positive to 5σ is comparable to, or even greater than, the integration time required to obtain a 10 nm spectrum with a signal-to-noise of 5.*

This is a counter-intuitive result. We would normally expect photometric characterization to take less telescope time than obtaining a spectrum. However, differences in color between Earth and the nearby false-positives are subtle, and so relatively long integration times are required to resolve this difference. Additionally, because the color-color space is defined by reflectivity ratios, with an observational uncertainty in both the numerator and denominator, error propagation dictates that individual photometric observations need to be quite precise to accurately place Earth in color-color space. This is reflected in the fact that the $S/N$ in each photometric bin needs to be 27.6 to ensure the color-color space separation signal-to-noise of 5.

If we repeat the detectability calculation but assume a separation of 0.5 between Earth and its nearest neighbor then we find $t_{spectrum} \approx 0.8 t_{color}$. This shows that even if we ignore all the possible icy false positives (i.e. Fig. 4), separating Earth from other false positives using optimal photometric bins still takes about the same amount of telescope time as obtaining a spectrum. It should also be noted that changing the minimum bin size does not affect this conclusion since, as demonstrated above in Fig. 8, optimal separation is only weakly dependent on minimum bin size. Additionally, if we consider the integration time required to separate Earth from all false positives to only 3σ, compared to the integration time required to obtain a 10 nm spectrum with S/N of 5, then we find $t_{spectrum} \approx 0.9 t_{color}$. For a separation of 2σ we find $t_{spectrum} \approx 2 t_{color}$.

For the first generation of coronagraph-equipped space telescopes capable of imaging habitable exoplanets (e.g. WFIRST-AFTA or Exo-C) the dominant noise source will be dark current in the detector (Robinson, et al. 2015). For this second scenario, we replace equation (1.6) with the following signal-to-noise expression:

$$S/N = \frac{F_{Planet}}{\sqrt{F_{Dark}}} \sqrt{t} \quad (1.10)$$

Here, $F_{Dark}$ is the dark current count rate. Robinson, et al. (2015) estimates that for an integral field spectrometer, the dark current flux in spectroscopic observations will be six times that for photometric observations. This is because the critical sampling of each spectral element over two pixels in the dispersion direction on the CCD is multiplied by the perpendicular spread of the spectrum across roughly three pixels in width. We therefore have:

$$\begin{aligned} t_{spectrum} &= \left( \frac{(S/N)_{spectrum} \sqrt{F_{Dark-spectrum}} / F_{Planet}'}{(S/N)_{color} \sqrt{F_{Dark-color}} / F_{Planet}} \right)^2 t_{color} \\ &= \left( \frac{5\sqrt{6 F_{Dark-color}} / F_{Planet}'}{27.6 \sqrt{F_{Dark-color}} / 10 F_{Planet}'} \right)^2 t_{color} \\ &= 6 \left( \frac{50}{27.6} \right)^2 t_{color} \approx 20 t_{color} \end{aligned} \quad (1.11)$$

Thus, *if observational noise is dominated by dark current then the integration time required to separate Earth from the nearest false positive to 5σ is twenty times less than the integration time required to obtain a 10 nm spectrum with a signal-to-noise of 5.* This indicates that photometry could be practically applied to identify probable Earth-like planets for spectroscopic follow-up if observational noise is dominated by dark current.

## 4. DISCUSSION

Although the set of uninhabitable false positive planets considered in this study is extensive, it is not intended to be exhaustive. Exotic surface mineralogy or other pathological scenarios could potentially produce spectra that mimic Earth's apparently unique spectrum. For example, a growing literature investigates hypothetical scenarios whereby atmospheric oxygen could accumulate in the absence of life. If these false-positive scenarios are realistic, then some planets may occupy a similar region in color-color space to Earth. Kasting, et al. (1993) argued that a runaway greenhouse on a planet interior to the habitable zone, or the photochemical production of oxygen on a frozen Mars-like planet exterior to the habitable zone, could result in detectable levels of oxygen. These false-positives could eventually be distinguished on the basis of semi-major axis and the absence of water vapor, but a photometric survey may not discriminate these variables. Wordsworth and Pierrehumbert (2014) and Luger and Barnes (2015) outline scenarios whereby water loss on otherwise habitable planets could result in the accumulation of atmospheric oxygen. Once again, these scenarios could eventually be distinguished through a lack of non-condensable gases and subtle pressure-sensitive spectral features (Misra et al. 2014; Schwieterman et al. 2015b), respectively, but photometry alone could not differentiate these false positives. Domagal-Goldman et al. (2014), Tian et al. (2014) and Harman et al. (2015) also investigated oxygen false positives within the habitable zone. Harman, et al. (2015) concluded that it is extremely difficult to produce false positive oxygen around G and F stars through photochemistry alone, but that abiotic oxygen could accumulate around K and M stars if surface sinks, such as reactions with dissolved ferrous iron or CO, are small. In short, the variety and plausibility of false-positive oxygen accumulation is still under discussion. For this reason, planet color is not a means of definitive characterization, but rather an indicator that certain objects are sufficiently interesting to justify more extensive observations. In this way, planet color could be incorporated into the observing strategy for future direct imaging missions (see below).

It could be asked whether the false-positive model planets used in this study – especially the icy worlds that are most similar in color to Earth – are realistic objects. Firstly we consider the Mars-like false positives: planets with a Mars-like surface and a thick $CO_2$ or $N_2$ atmosphere. There is no strong reason to believe that such objects are unlikely to exist. In fact, it has been argued that Mars may have had a thick $CO_2$ atmosphere early in its history due to greater levels of volcanic outgassing (e.g. review by Haberle et al. 2016).

It is less clear whether exoplanets with an icy surface like that of Callisto or Europa could form with a thick, pure $N_2$ atmosphere. The only example of an icy body with a thick $N_2$ atmosphere in the Solar System is Titan, and in this case the presence of atmospheric methane causes an optical haze that completely changes the disk-averaged reflectance spectrum. Titan's atmospheric $CH_4$ is completely destroyed by photodissociation on a timescale of 30-100 m.y., and so either it is being replenished by a subsurface reservoir or we are observing Titan at a special time in its history (Lunine et al. 1989). In the latter case, there would have been times in Titan's history when its atmosphere was more similar to our icy false positives. Even if it is not possible for icy bodies to form with thick, pure $N_2$ atmospheres, we can envisage a scenario whereby a terrestrial planet with an $N_2$ atmosphere formed interior to the snow line, then later migrated outwards. Since the planet would be too warm to preserve subsurface methane clathrates in its original location, post-migration there would be no subsurface reservoir to replenish $CH_4$ lost to photodissociation and subsequent hydrogen escape. Additionally, any liquid water would freeze out post-migration resulting in an icy surface. Thus, it is possible to imagine an icy body with a relatively thick, pure $N_2$ atmosphere after atmospheric $CH_4$ has been depleted. Such a planet might be observed to reside outside the traditional habitable zone (through time-dependent photometry or existing transit or radial velocity data), but the potential photometric similarity to modern Earth cautions us against permissive interpretations of planet colors in the extended habitable zones that have recently been proposed (Pierrehumbert & Gaidos 2011; Seager 2013).

Although we have treated the Earth and the Moon separately in this analysis, direct imaging missions will not separately resolve exo-Earths from companion satellites, and will instead obtain a combined spectrum. This could have an effect on an exo-Earth's position in color-color space. Robinson (2011) showed that in the infrared the presence of the Moon has a significant effect on the combined Earth-Moon emission spectrum. However, in the visible, the presence of the Moon should not have a large effect on the Earth's color. The radii and average albedos for the Earth and The Moon are 6400 km and 0.3, and 1700 km and 0.1, respectively. This implies that at full phase the radiance of The Moon is approximately $(1700^2 \times 0.1)/(6400^2 \times 0.3) = 2\%$ that of Earth. In addition, the Moon's spectrum is relatively flat and featureless, and so we would not expect the Moon to have a large effect on the combined Earth-Moon color. Of course, exo-Earths with larger, higher-albedo companions could be considerably different in color, and so the possibility for false negatives cannot be completely ignored. However, although the exoplanet-exomoon system is unresolved, if the time- and wavelength-dependent positional variation of the center of light of the combined exoplanet-exomoon system can be measured, then the exoplanet and exomoon spectrum could be disentangled (Agol et al. 2015).

*4.1 Implications for observing strategies and telescope design*

There are numerous scenarios whereby uninhabitable exoplanets could appear very similar in color to Earth. Even with optimized photometric bins the distinction between Earth and these uninhabitable exoplanets is subtle. For 2-meter class direct imaging space telescopes such as WFIRST-AFTA, Exo-C or Exo-S, exoplanet observations will mostly be dominated by dark noise (Robinson, et al. 2015). In this scenario, photometric identification of exo-Earths will require much shorter integration times than spectroscopic characterization. However, for these 2-meter class telescopes it will only be possible to image potentially habitable terrestrial planets around a small number of nearby stars, and so doing a photometric survey of habitable planets will not be possible.

For the next generation of large-aperture (~10 m) space telescope such as ATLAST/LUVOIR, the utility of photometric surveys will depend strongly on dark current. If dark current dominates the observational noise, then photometric characterization could be readily incorporated into an observing strategy, such as that for ATLAST/LUVOIR described by Domagal-Goldman et al. (2015). For example, after preliminary observations have characterized the system architecture and determined if any planets are in the habitable zone, the next step would be to obtain broadband photometry (100 nm bins) of any potentially habitable planets. The optimal bins from Fig. 2, 3 and 7 (431-531 nm, 569-693 nm, 770-894) could be used to identify probable exo-Earths for spectroscopic follow-up. In this scenario optimized photometry would be a valuable tool for preliminary characterization.

Alternatively, if astrophysical noise dominates ATLAST/LUVOIR observations, then the integration times required to separate Earth from uninhabitable positives using optimal photometric bins are comparable to the integration times required to obtain a spectrum. Thus there is little advantage in using visible wavelength photometry first, before spectroscopy, to identify potential Earth-like planets for spectroscopic follow-up. Instead, color could be calculated from a low-resolution spectrum and used as an additional piece of information to evaluate whether higher-resolution spectral observations are justified. For example, after preliminary observations have characterized the system the next step would be to obtain low-resolution spectra (e.g. 50 nm bins, S/N=5) of any potentially habitable planets. These observations would be insufficient to photometrically separate Earth from false positives unambiguously – in fact, they would separate Earth from its nearest neighbor by only 1.3σ in color-color space. However, this crude color information still constrains the region in color-color space the exoplanet occupies, and hence it could be used in combination with suggestive absorption features in the spectrum (such as water absorption) to evaluate whether a high resolution spectrum should be obtained, or whether the telescope should move on to its next target. It may be preferable to use the optimal photometric bins in Fig. 4 for this purpose (350-450 nm, 593-693 nm, and 879-992 nm). This is because the purpose of these photometric observations is not to definitively identify exo-Earths, but rather identify potentially Earth-like planets for higher-resolution spectroscopic follow-up. The Fig. 4 bins have the advantage of better-separating Earth from all false positives except the icy bodies with thick atmospheres, which is appropriate given the coarseness of the spectrum (there will be large uncertainties in color-color position). Additionally, it may be possible to distinguish icy bodies from Earth-like planets with a low-resolution spectrum from the presence or absence of water vapor absorption.

*4.2 Optimal wavelength range*

In the original Earth-separating optimization with all false positives included (Fig. 2) we found that despite allowing the optimization to consider all photometric bins between 350 nm and 1000 nm, the solution only uses bins between 431 nm and 894 nm. This suggests that extending the wavelength capabilities of a direct imaging telescope beyond this range is not necessary to uniquely identify potential Earth-like planets with photometry (note however that extending wavelength to 1000 nm is highly desirable for spectroscopy because it provides access to the 940 nm water vapor band, which is critical for assessing habitability).

This conclusion does not hold if we accept the possibility of icy false positives (Fig. 4). In this case, if the wavelength range available for photometric bins is made smaller, then the optimal bins do not separate Earth as clearly as the 350-1000 nm case ($\lambda_{min} = 350, \lambda_{max} = 1000$). For instance, if the photometric bins are restricted to 450-1000 nm range then the Earth is only separated from its neighbor by a distance of 0.34 in color-color space (not shown), compared to 0.53 for the full wavelength range in Fig. 4. The separation is poor because the restricted wavelength range does not fully capture Earth's exceptional shortwave reflectivity due to Rayleigh scattering. In the absence of icy false positives, shortwave reflectivity is the most diagnostic feature of Earth's spectrum. If an intermediate wavelength range is applied ($\lambda_{min} = 400, \lambda_{max} = 1000$) then the separation between Earth and its nearest neighbor is 0.47. This suggests that telescopes with a minimum wavelength of 400 nm receive most of the benefits of short wavelength photometric characterization. Currently, the direct imaging mission proposals Exo-C and Exo-S are planned to operate from 450-1000 nm and 400-1000 nm respectively (Seager, et al. 2015; Stapelfeldt, et al. 2015), whilst the WFIRST-AFTA coronograph will operate between 430-970 nm with spectrometry restricted to 600-970 nm (Spergel, et al. 2015). The HDST concept has a minimum wavelength range of 100-2000 nm (Dalcanton, et al. 2015), whereas the similar ATLAST concept has a 110-2500 nm wavelength range (Postman, et al. 2009).

We have not explored the utility of photometry shortward of 350 nm. This is because the model spectra we generated are unreliable in the UV due to incomplete line lists and absorption cross sections. However, UV photometry could be useful for identifying exo-Earths due to strong ozone absorption at λ<320 nm. This has been previously noted by Heap et al. (2008), who argued that Hartley band (ozone absorption) is a promising biosignature in the UV. However, it should be noted that recent photochemical modeling has shown that $O_3$ is much easier to produce abiotically in a terrestrial planet atmosphere than $O_2$ (Domagal-Goldman, et al. 2014; Harman, et al. 2015), thus complicating the interpretation of an $O_3$ signature. An upper wavelength limit of 1000 nm was used in this study because this is the upper wavelength limit for near-term direct imaging missions, and because the spectra available in the literature are typically truncated at around 1000 nm. However, the utility of photometry longward of 1000 nm could also be explored in future work.

## 5. CONCLUSIONS

- Icy bodies with thick $N_2$ atmospheres or Mars-like surfaces with thick $N_2$ or $CO_2$ atmospheres are very similar in color to the Earth. The thick atmospheres of these uninhabitable, quasi pale blue dots mimic the blue reflectance of Earth's Rayleigh scattering.

- However, the U-shape of Earth's spectrum from the far-UV to the near IR is unlike any known uninhabitable planet or our model uninhabitable planets, and at least partly biogenic in origin. The optimal photometric bins for separating Earth from uninhabitable planets exploit this feature of Earth's spectrum. The wavelength ranges of these bins are 431-531 nm, 569-693 nm, 770-894 nm (Fig. 2, 3 and 7).

- The utility of broadband photometry for identifying exo-Earths depends strongly on the primary source of observational noise. If observations are photon limited or dominated by astrophysical noise (e.g. zodiacal light or imperfect starlight suppression) then the telescope integration time required to identify Earth's unique color is comparable to - or even greater than - the integration time required to obtain a moderate resolution (R~70) spectrum. In this scenario there is little advantage to using visible wavelength photometry first, before spectroscopy, to identify potential Earth-like planets for spectroscopic follow-up.

- However, if observational noise is dominated by dark current comparable to those of present detectors, then the telescope integration time required to identify Earth's unique color is ~20 times shorter than the integration time required to obtain a moderate resolution (R~70) spectrum. In this scenario optimized photometric bins could be used to identify probable exo-Earths for spectroscopic follow-up.

- Extrasolar Archean Earths are extremely difficult to uniquely identify using visible broadband photometry. Additionally, the unique, biogenic shape of Earth's spectrum has only been present during the Phanerozoic (0.5-0 Ga). This implies that for the majority of its history Earth would be challenging to uniquely identify using photometry alone.

**ACKNOWLEDGEMENTS**

We thank Shawn Domagal-Goldman, Aomawa Shields, Elena Amador, the anonymous reviewer, and the ATLAST study team for helpful feedback and discussions. This work was performed as part of the NASA Astrobiology Institute's Virtual Planetary Laboratory, supported by the National Aeronautics and Space Administration through the NASA Astrobiology Institute under solicitation NNH12ZDA002C and Cooperative Agreement Number NNA13AA93A. BC and TDR are supported by the NAI in the NASA Postdoctoral Program. Discussions with Shawn Domagal-Goldman and the ATLAST study team were enabled by a NASA Astrobiology Early Career Collaboration Award. This work benefited from the use of advanced computational, storage, and networking infrastructure provided by the Hyak supercomputer system at the University of Washington.

**REFERENCES**

Agol, E., Jansen, T., Lacy, B., Robinson, T. D., & Meadows, V. 2015, The Astrophysical Journal, 812, 5

Andrews, D. G. 2010, An introduction to atmospheric physics (Cambridge University Press)

Arney, G., Meadows, V., Domagal-Goldman, S., Claire, M., & Schwieterman, E. 2015. in American Astronomical Society Meeting Abstracts, Hazy Archean Earth as an Analog for Hazy Earthlike Exoplanets

Arnold, L., Gillet, S., Lardière, O., Riaud, P., & Schneider, J. 2002, Astronomy & Astrophysics, 392, 231

Blandford, R., Hillenbrand, L., Haynes, M. P., Huchra, J. P., Rieke, M. J., & others. 2010, in Decadal Survey of Astronomy and Astrophysics 2010, "New Worlds, New Horizons in Astronomy & Astrophysics" National Research Council, National Academy Press, ISBN-10: 0-309-15799-4

Blaney, D. L., Walsh, P. A., & McCord, T. B. 1988, in In Lunar and Planetary Inst, MEVTV Workshop on Nature and Composition of Surface Units on Mars p 25-27 (SEE N88-29654 23-91)

Botet, R., Rannou, P., & Cabane, M. 1997, Applied optics, 36, 8791

Cahoy, K. L., Marley, M. S., & Fortney, J. J. 2010, The Astrophysical Journal, 724, 189

Cash, W. 2006, Nature, 442, 51

Catling, D. C. 2015, in Treatise on Geophysics (2nd Ed), ed. G. Schubert (New York: Elsevier), 429

Charnay, B., Meadows, V., & Leconte, J. 2015, submitted to ApJ

Claire, M. W., Sheets, J., Cohen, M., Ribas, I., Meadows, V. S., & Catling, D. C. 2012, The Astrophysical Journal, 757, 95

Clark, R. N., Swayze, G. A., Wise, R., Livo, K. E., Hoefen, T. M., Kokaly, R. F., & Sutley, S. J. 2007, in (US Geological Survey Reston, VA)

Cockell, C. S., et al. 2009, Astrobiology, 9, 1

Cowan, N. B., et al. 2011, The Astrophysical Journal, 731, 76

Crisp, D. 1990, Journal of Geophysical Research: Solid Earth (1978–2012), 95, 14577

Crisp, D. 1997, Geophysical Research Letters, 24, 571

Crow, C. A., et al. 2011, The Astrophysical Journal, 729, 130

Dalcanton, J., et al. 2015, arXiv preprint arXiv:150704779

Delorme, P., et al. 2013, Astronomy & Astrophysics, 553, L5

Des Marais, D. J., et al. 2002, Astrobiology, 2, 153

Domagal-Goldman, S., et al. 2015, in Astrobiology Science Conference (Chicago)

Domagal-Goldman, S. D., Kasting, J. F., Johnston, D. T., & Farquhar, J. 2008, Earth and Planetary Science Letters, 269, 29

Domagal-Goldman, S. D., Segura, A., Claire, M. W., Robinson, T. D., & Meadows, V. S. 2014, The Astrophysical Journal, 792, 90

Driese, S. G., Jirsa, M. A., Ren, M., Brantley, S. L., Sheldon, N. D., Parker, D., & Schmitz, M. 2011, Precambrian Research, 189, 1

Evans, T. M., et al. 2013, The Astrophysical Journal Letters, 772, L16

Fressin, F., et al. 2013, The Astrophysical Journal, 766, 81

Grenfell, T. C., Warren, S. G., & Mullen, P. C. 1994, Journal of Geophysical Research: Atmospheres (1984–2012), 99, 18669

Guyon, O., Pluzhnik, E. A., Galicher, R., Martinache, F., Ridgway, S. T., & Woodruff, R. A. 2005, The Astrophysical Journal, 622, 744

Haberle, R. M., Catling, D. C., Carr, M. H., & Zahnle, K. 2016, in The Atmosphere and Climate of Mars, ed. R. M. e. a. Haberle (Cambridge Univ. Press, in press.)

Hamdani, S., et al. 2006, Astronomy & Astrophysics, 460, 617

Haqq-Misra, J. D., Domagal-Goldman, S. D., Kasting, P. J., & Kasting, J. F. 2008, Astrobiology, 8, 1127

Harman, C. E., Schwieterman, E. W., Schottelkotte, J. C., & Kasting, J. F. 2015, The Astrophysical Journal, 812:137
20

Heap, S. R., Lindler, D., & Lyon, R. 2008. in SPIE Astronomical Telescopes+ Instrumentation, Detecting biomarkers in exoplanetary atmospheres with a Terrestrial Planet Finder (International Society for Optics and Photonics), 70101N
Hegde, S., & Kaltenegger, L. 2013, Astrobiology, 13, 47
Hegde, S., Paulino-Lima, I. G., Kent, R., Kaltenegger, L., & Rothschild, L. 2015, Proceedings of the National Academy of Sciences, 112, 3886
Howard, A. W., et al. 2010, Science, 330, 653
Hulburt, E. 1953, JOSA, 43, 113
Janson, M., et al. 2013, The Astrophysical Journal Letters, 778, L4
Kaltenegger, L., et al. 2010, Astrobiology, 10, 89
Kaltenegger, L., Traub, W. A., & Jucks, K. W. 2007, The Astrophysical Journal, 658, 598
Karkoschka, E. 1994, Icarus, 111, 174
---. 1998, Icarus, 133, 134
Kasper, M., et al. 2010. in SPIE Astronomical Telescopes+ Instrumentation, EPICS: direct imaging of exoplanets with the E-ELT (International Society for Optics and Photonics), 77352E
Kasting, J. F., & Ackerman, T. P. 1986, Science, 234, 1383
Kasting, J. F., Liu, S. C., & Donahue, T. M. 1979, Journal of Geophysical Research: Oceans, 84, 3097
Kasting, J. F., Whitmire, D. P., & Reynolds, R. T. 1993, Icarus, 101, 108
Khare, B. N., Sagan, C., Arakawa, E., Suits, F., Callcott, T., & Williams, M. 1984, Icarus, 60, 127
Kiang, N. Y., et al. 2007, Astrobiology, 7, 252
Kopparapu, R. K., et al. 2013, The Astrophysical Journal, 765, 131
Kuzuhara, M., et al. 2013, The Astrophysical Journal, 774, 11
Levy, H. 1971, Science, 173, 141
Livengood, T. A., et al. 2011, Astrobiology, 11, 907
Lodders, K. 2003, The Astrophysical Journal, 591, 1220
Lucey, P. G., Taylor, G. J., & Malaret, E. 1995, Science, 268, 1150
Luger, R., & Barnes, R. 2015, Astrobiology, 15, 119
Lunine, J., Atreya, S., & Pollack, J. 1989, Origin and evolution of planetary and satellite atmospheres, 1, 605
Macintosh, B., et al. 2015, Science, 350, 64
Macintosh, B., et al. 2014, Proceedings of the National Academy of Sciences, 111, 12661
Marcy, G. W., Weiss, L. M., Petigura, E. A., Isaacson, H., Howard, A. W., & Buchhave, L. A. 2014, Proceedings of the National Academy of Sciences, 111, 12655
Marois, C., et al. 2008, Science, 322, 1348
Matsuo, T., & Tamura, M. 2010. in SPIE Astronomical Telescopes+ Instrumentation, Second-earth imager for TMT (SEIT): a proposal and concept Description (International Society for Optics and Photonics), 773584
McCord, T. B., Clark, R. N., & Singer, R. B. 1982, Journal of Geophysical Research, 87, 3021
Meadows, V. S. 2006, Proceedings of the International Astronomical Union, 1, 25
Meadows, V. S., & Crisp, D. 1996, Journal of Geophysical Research: Planets (1991–2012), 101, 4595
Misra, A., Meadows, V., Claire, M., & Crisp, D. 2014, Astrobiology, 14, 67
NASA. 2013, Cassini images PIA14949 and PIA17170, JPL Photojournal
Pavlov, A. A., Brown, L. L., & Kasting, J. F. 2001, Journal of Geophysical Research: Planets (1991–2012), 106, 23267
Pierrehumbert, R., & Gaidos, E. 2011, The Astrophysical Journal Letters, 734, L13
Postman, M., et al. 2009, arXiv preprint arXiv:09040941
Rameau, J., et al. 2013, The Astrophysical Journal Letters, 772, L15
Robinson, T. D. 2011, The Astrophysical Journal, 741, 51

Robinson, T. D., Ennico, K., Meadows, V. S., Sparks, W., Bussey, D. B. J., Schwieterman, E. W., & Breiner, J. 2014, The Astrophysical Journal, 787, 171
Robinson, T. D., Meadows, V. S., & Crisp, D. 2010, The Astrophysical Journal Letters, 721, L67
Robinson, T. D., et al. 2011, Astrobiology, 11, 393
Robinson, T. D., Stapelfeldt, K. R., & Marley, M. S. 2015, PSAP, in press
Rothman, L., Gordon, I., Babikov, Y., Barbe, A., Benner, D. C., & Bernath, P. 2013, J Quant Spectrosc Radiat Transfer
Rothman, L. S., et al. 2009, Journal of Quantitative Spectroscopy and Radiative Transfer, 110, 533
Rowe, J. F., et al. 2006, The Astrophysical Journal, 646, 1241
Sagan, C. 1994, Pale blue dot: A vision of the human future in space (Random House)
Sagan, C., Thompson, W. R., Carlson, R., Gurnett, D., & Hord, C. 1993, Nature, 365, 715
Sanromá, E., Pallé, E., & Munõz, A. G. 2013, The Astrophysical Journal, 766, 133
Sanromá, E., Pallé, E., Parenteau, M., Kiang, N., Gutiérrez-Navarro, A., López, R., & Montañés-Rodríguez, P. 2014, The Astrophysical Journal, 780, 52
Schwieterman, E., Cockell, C., & Meadows, V. 2015a, Astrobiology, 15, 341
Schwieterman, E. W., Robinson, T. D., Meadows, V. S., Misra, A., & Domagal-Goldman, S. 2015b, The Astrophysical Journal, 810, 57
Seager, S. 2013, Science, 340, 577
Seager, S., et al. 2015, in (Pasadena, California: Jet Propulsion Laboratory)
Seager, S., Schrenk, M., & Bains, W. 2012, Astrobiology, 12, 61
Seager, S., Turner, E. L., Schafer, J., & Ford, E. B. 2005, Astrobiology, 5, 372
Segura, A., et al. 2005, Astrobiology, 5, 706
Segura, A., et al. 2003, Astrobiology, 3, 689
Shields, A. L., Meadows, V. S., Bitz, C. M., Pierrehumbert, R. T., Joshi, M. M., & Robinson, T. D. 2013, Astrobiology, 13, 715
Spergel, D., et al. 2015, arXiv preprint arXiv:150303757
Spiegel, D. S., Burrows, A., Ibgui, L., Hubeny, I., & Milsom, J. A. 2010, The Astrophysical Journal, 709, 149
Stapelfeldt, K., et al. 2015, in (Pasadena, California: Jet Propulsion Laboratory)
Sudarsky, D., Burrows, A., Hubeny, I., & Li, A. 2005, The Astrophysical Journal, 627, 520
Sudarsky, D., Burrows, A., & Pinto, P. 2000, The Astrophysical Journal, 538, 885
Tian, F., France, K., Linsky, J. L., Mauas, P. J., & Vieytes, M. C. 2014, Earth and Planetary Science Letters, 385, 22
Tinetti, G., Meadows, V. S., Crisp, D., Fong, W., Velusamy, T., & Snively, H. 2005, Astrobiology, 5, 461
Tinetti, G., et al. 2006, Astrobiology, 6, 881
Tolfo, F. 1977, Journal of Aerosol Science, 8, 9
Traub, W. A. 2003a. in Scientific Frontiers in Research on Extrasolar Planets, The colors of extrasolar planets, 595
---. 2003b. in Earths: DARWIN/TPF and the Search for Extrasolar Terrestrial Planets, Extrasolar planet characteristics in the visible wavelength range, eds. M. Fridlund, & T. Henning (Heidelberg, Germany: ESA Publications Division), 231
Warren, S. G., Brandt, R. E., Grenfell, T. C., & McKay, C. P. 2002, Journal of Geophysical Research: Oceans (1978–2012), 107, 31
Wordsworth, R., & Pierrehumbert, R. 2014, The Astrophysical Journal Letters, 785, L20
Wordsworth, R. D., Forget, F., Selsis, F., Millour, E., Charnay, B., & Madeleine, J.-B. 2011, The Astrophysical Journal Letters, 733, L48
Yan, F., et al. 2015, International Journal of Astrobiology, 14, 255
Zerkle, A. L., Claire, M. W., Domagal-Goldman, S. D., Farquhar, J., & Poulton, S. W. 2012, Nature Geoscience, 5, 359

**TABLES**

Table 1: This table lists all the planet spectra used in this study. The bolded rows represent the set of uninhabitable false positives, and the italicized rows are the subset of icy false positives referred to in the main text. The color-color plot symbols for each planet are specified. Note that these are consistent across all figures in this study. The source of each spectrum – either a citation or SMART model output – is also listed. All spectra are observed at quadrature unless otherwise stated.

| Planet/Moon | Notes | Symbol | Source for spectrum |
|---|---|---|---|
| Earth quadrature phase | This Earth spectrum was used in all optimization calculations. | Blue circle | Validated 3D Earth model and SMART |
| Earth full phase | | Blue diamond | Validated 3D Earth model and SMART |
| Earth gibbous phase | | Blue star | Validated 3D Earth model and SMART |
| Earth crescent phase | | Blue downward triangle | Validated 3D Earth model and SMART |
| **The Moon** | | **Black circle** | **Validated 1D model and SMART** |
| **Mars** | | **Red circle** | **Validated 1D model and SMART** |
| **Venus** | | **Black circle** | **Validated 1D model and SMART** |
| **Jupiter** | **Full phase** | **Black circle** | *Karkoschka (1998)* |
| **Saturn** | **Full phase** | **Black circle** | *Karkoschka (1998)* |
| **Uranus** | **Full phase** | **Black circle** | *Karkoschka (1998)* |
| **Neptune** | **Full phase** | **Black circle** | *Karkoschka (1998)* |
| **Titan** | **Full phase** | **Black circle** | *Karkoschka (1998)* |
| **Io** | **Full phase** | **White circle** | *Karkoschka (1994)* |
| **Callisto** | **Full phase** | **Cyan circle** | *Karkoschka (1994)* |
| **Ganymede** | **Full phase** | **Purple circle** | *Karkoschka (1994)* |
| **Europa** | **Full phase** | **Green circle** | *Karkoschka (1994)* |
| **Giant planet class I** | **Full phase** | **Black circle** | *Sudarsky, et al. (2000)* |
| **Giant planet class II** | **Full phase** | **Black circle** | *Sudarsky, et al. (2000)* |
| **Giant type I with haze** | **Full phase** | **Black circle** | *Sudarsky, et al. (2000)* |
| **Sub-Neptune thick $H_2$ atmosphere** | **6.6 $m_\oplus$ and 2.4 $R_\oplus$ with 2000 W/m² insolation and solar metallicity** | **Black circle** | **SMART** |
| **Sub-Neptune thick $H_2$ atmosphere high** | **6.6 $m_\oplus$ and 2.4 $R_\oplus$ with** | **Black circle** | **SMART** |
| metallicity | 2000 W/m² insolation 100x and solar metallicity | | |
| Sub-Neptune thick H₂O atmosphere | 6.6 $m_\oplus$ and 2.4 $R_\oplus$ with 2000 W/m² insolation | Black circle | SMART |
| Snow | 100% snow-covered planet with clear sky Earth-like atmosphere | Black circle | *SMART, surface spectrum from Shields, et al. (2013)* |
| Blue ice | 100% blue ice-covered planet with clear sky Earth-like atmosphere | Black circle | *SMART, surface spectrum from Shields, et al. (2013)* |
| Kaolinite | 100% kaolinite covered planet with clear sky Earth-like atmosphere | Black circle | *SMART, surface spectrum from Clark et al. (2007)* |
| *Callisto 0.5 bar N₂* | *250 K isothermal atmosphere.* | *Cyan upward triangle* | *SMART, surface spectrum from Karkoschka (1994)* |
| *Callisto 0.75 bar N₂* | " | *Cyan star* | *SMART, surface spectrum from Karkoschka (1994)* |
| *Callisto 1 bar N₂* | " | *Cyan square* | *SMART, surface spectrum from Karkoschka (1994)* |
| *Callisto 2 bar N₂* | " | *Cyan downward triangle* | *SMART, surface spectrum from Karkoschka (1994)* |
| *Europa 0.5 bar N₂* | " | *Green upward triangle* | *SMART, surface spectrum from Karkoschka (1994)* |
| *Europa 1 bar N₂* | " | *Green square* | *SMART, surface spectrum from Karkoschka (1994)* |
| *Europa 2 bar N₂* | " | *Green downward triangle* | *SMART, surface spectrum from Karkoschka (1994)* |
| *Ganymede 0.5 bar N₂* | " | *Purple downward triangle* | *SMART, surface spectrum from Karkoschka (1994)* |
| *Ganymede 2 bar N₂* | " | *Purple square* | *SMART, surface spectrum from Karkoschka (1994)* |
| *Io 0.5 bar N₂* | " | *White square* | *SMART, surface spectrum from Karkoschka (1994)* |
| *Gray 0.2 bar N₂* | *Gray surface has albedo of 0.27. 250 K isothermal atmosphere.* | *Gray square* | *SMART* |
| *Gray 1.2 bar N₂* | " | *Gray downward triangle* | *SMART* |
| *Gray 2 bar N₂* | " | *Gray diamond* | *SMART* |
| *Mars 0.5 bar $N_2$* | *250 K isothermal atmosphere.* | *Red pentagon* | *SMART* |
| *Mars 2 bar $N_2$* | " | *Red leftward triangle* | *SMART* |
| *Mars 5 bar $N_2$* | " | *Red rightward triangle* | *SMART* |
| *Mars 0.5 bar $CO_2$* | " | *Red upward triangle* | *SMART* |
| *Mars 1 bar $CO_2$* | " | *Red square* | *SMART* |
| *Mars 1.5 bar $CO_2$* | " | *Red star* | *SMART* |
| *Mars 5 bar $CO_2$* | " | *Red downward triangle* | *SMART* |
| *Kaolinite 1 bar $CO_2$* | " | *Lime downward triangle* | *SMART, surface spectra from Clark, et al. (2007)* |
| *$CO_2$ ice 1 bar $N_2$* | *180 K isothermal atmosphere for consistency* | *Lime upward triangle* | *SMART, surface spectra from Tinetti et al. (2005)* |
| Archean Earth no haze, high (1E-2) $CH_4$ | 50% clear skies, 25% med cloud, 25% low cloud | Yellow circle | SMART |
| Archean Earth no haze, low (2E-3) $CH_4$ | " | Yellow star | SMART |
| Archean Earth with thick haze (1E-2 $CH_4$) | " | Yellow upward triangle | SMART |
| Archean Earth with thin haze (2E-3 $CH_4$) | " | Yellow leftward triangle | SMART |
| Earth with clear skies | Surface chosen to be linear mix of Ocean (65.6%), Kaolinite (5.5%), Grass (13.6%), Conifers (4%) and snow (11.3%) to match 24-hour averaged 3D model. | Blue pentagon | SMART |
| Earth 100% clear skies over ocean | | Blue upward triangle | SMART |
| Earth with 100% low cloud | | Blue rightward triangle | SMART |
| Earth with 100% mid cloud | | Blue leftward triangle | SMART |
| Earth with 100% high | | Blue square | SMART |

| cloud | | | |
| Jovian phase=0 | 3x solar metallicity with atmosphere in chemical equilibrium at 0.8 AU | Green star | Cahoy, et al. (2010) |
| Jovian phase=70 | " | Green star | Cahoy, et al. (2010) |
| Jovian phase=90 | " | Green star | Cahoy, et al. (2010) |
| Jovian phase=120 | " | Green star | Cahoy, et al. (2010) |
| Jovian phase=140 | " | Green star | Cahoy, et al. (2010) |
| Jovian phase=150 | " | Green star | Cahoy, et al. (2010) |

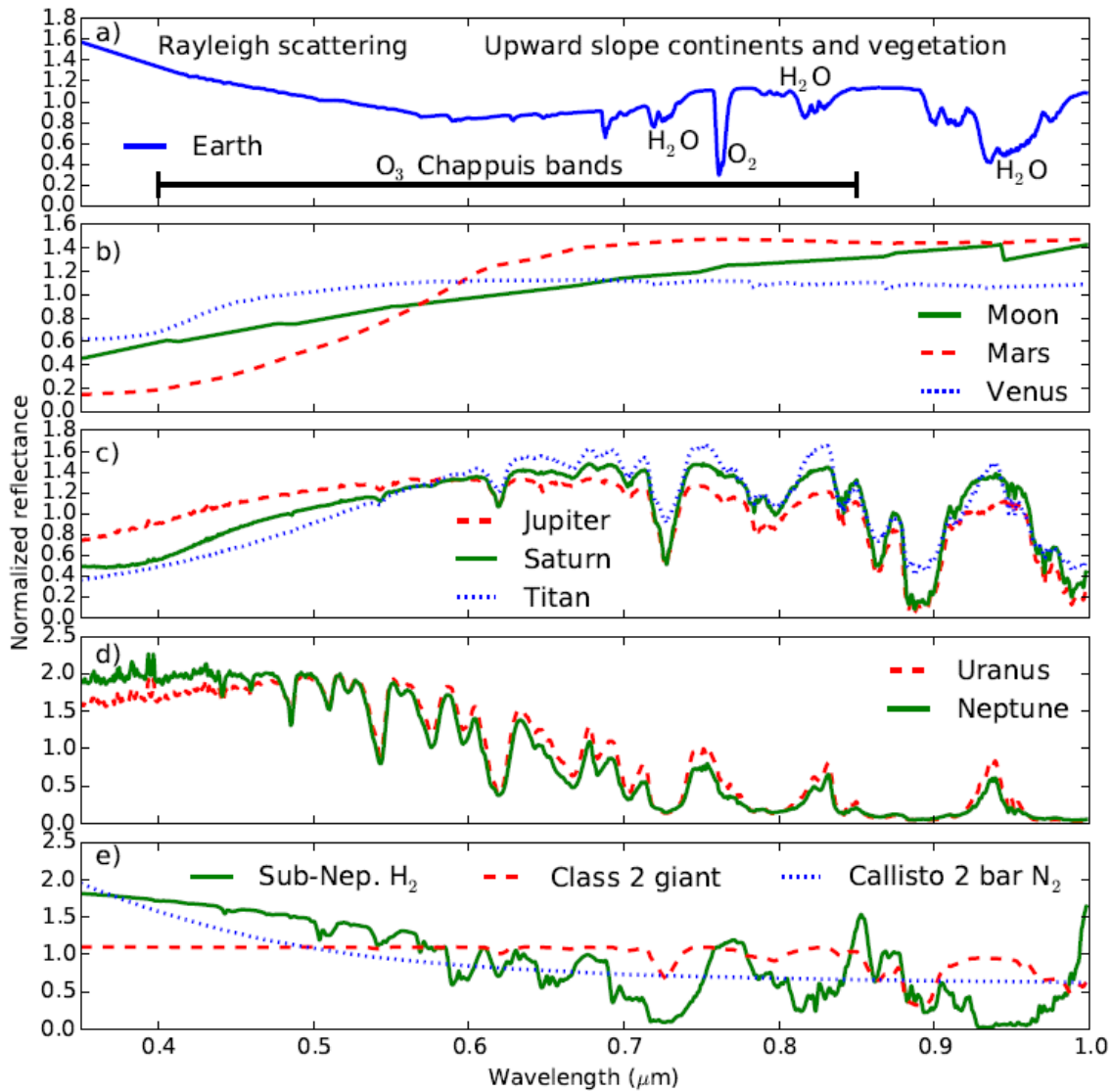

Figure 1: Reflectance spectra of Solar System objects and exoplanet models. (a) Earth is a pale blue dot due to strong Rayleigh scattering at short wavelengths. The spectrum is upward sloping from 600-900 nm due to continental reflectivity and the vegetation red edge. This, in combination with ozone absorption at 600 nm, gives the spectrum an overall U-shape. The black bar shows the width of the Chappuis ozone absorption bands (Andrews 2010). In contrast, objects with little or no atmosphere and a basaltic crust (Mars and the Moon in (b)) have an upward sloping spectrum. (c) Shows how Rayleigh scattering is suppressed in planets with optical hazes (see also Venus in (b)). (d) Demonstrates that Uranus and Neptune have qualitatively different spectra to Earth despite also being highly reflective in the blue. (e) Shows a selection of model exoplanet spectra including a sub Neptune, a giant planet, and an icy planet with a thick $N_2$ atmosphere. The references and descriptions for these spectra are described in full in the main text.

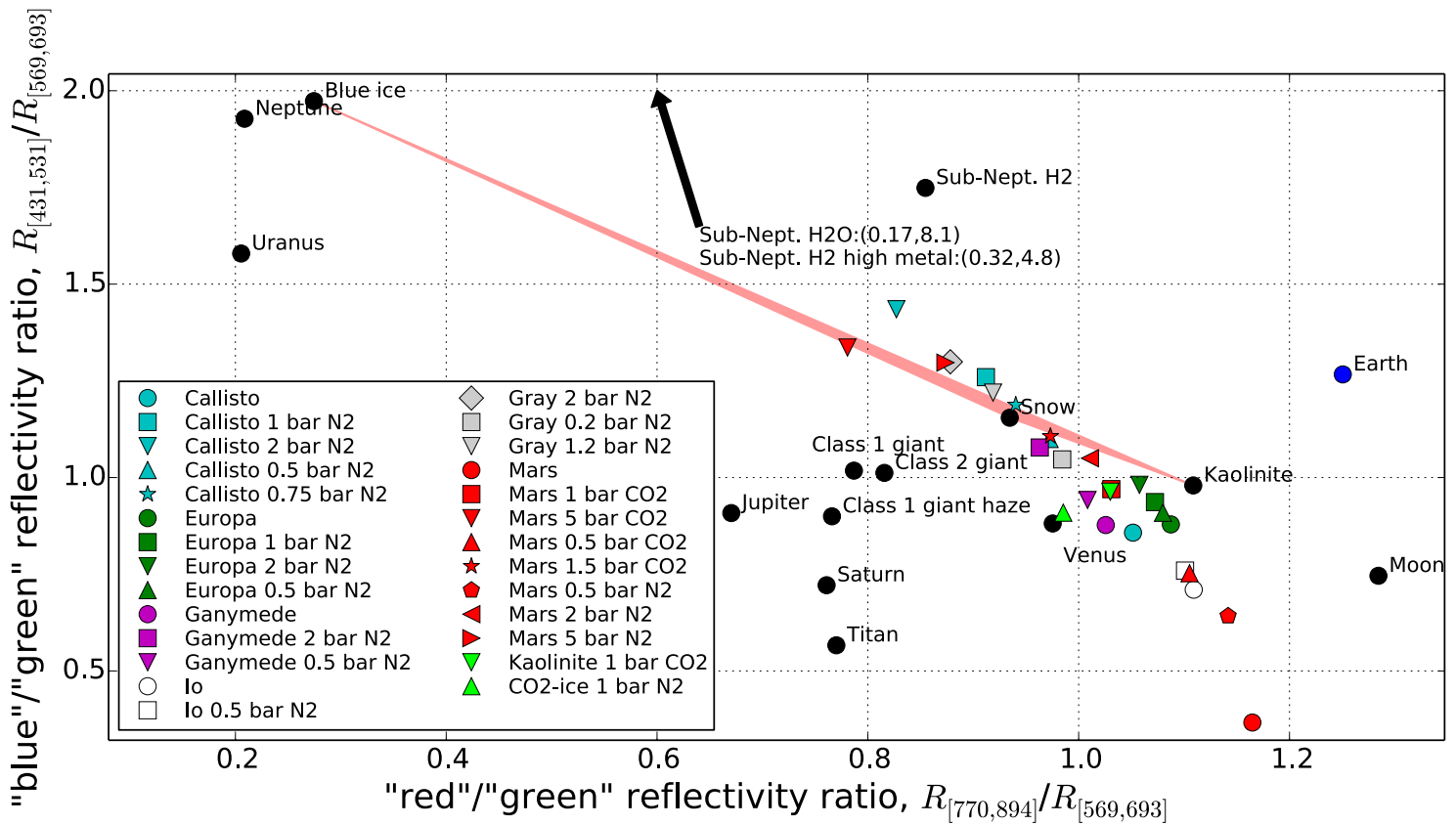

Figure 2: The Earth (blue circle) and all the uninhabitable planets (other symbols) – including both Solar System objects and exoplanet models – are plotted in color-color space, optimized to separate the Earth from its nearest neighbor. The horizontal axis is the mean reflectivity in the "red" bin (770-894 nm) divided by mean reflectivity in the "green" bin (569-693 nm), and the vertical axis is mean reflectivity in the "blue" bin (431-531 nm) divided by mean reflectivity in the "green" bin (569-693 nm). Earth's nearest neighbor is a Callisto with a 0.75 bar $N_2$ atmosphere, separated by 0.32 units (dimensionless). Note that the scaling of the horizontal axis is bigger than the vertical axis. The arrow denotes two sub-Neptune models that fall beyond the axis limits of this figure, and the red shaded region represents the range of colors possible for a snowball Earth planet (see main text). The Earth occupies a unique region in color-color space because of the U-shape of its reflectance spectrum: it is unusually blue for its level of redness and so both the "red"/"green" and "blue"/"green" reflectivities are appreciably greater than 1. The set of uninhabitable planets in this figure is described in full the main text and in Table 1.

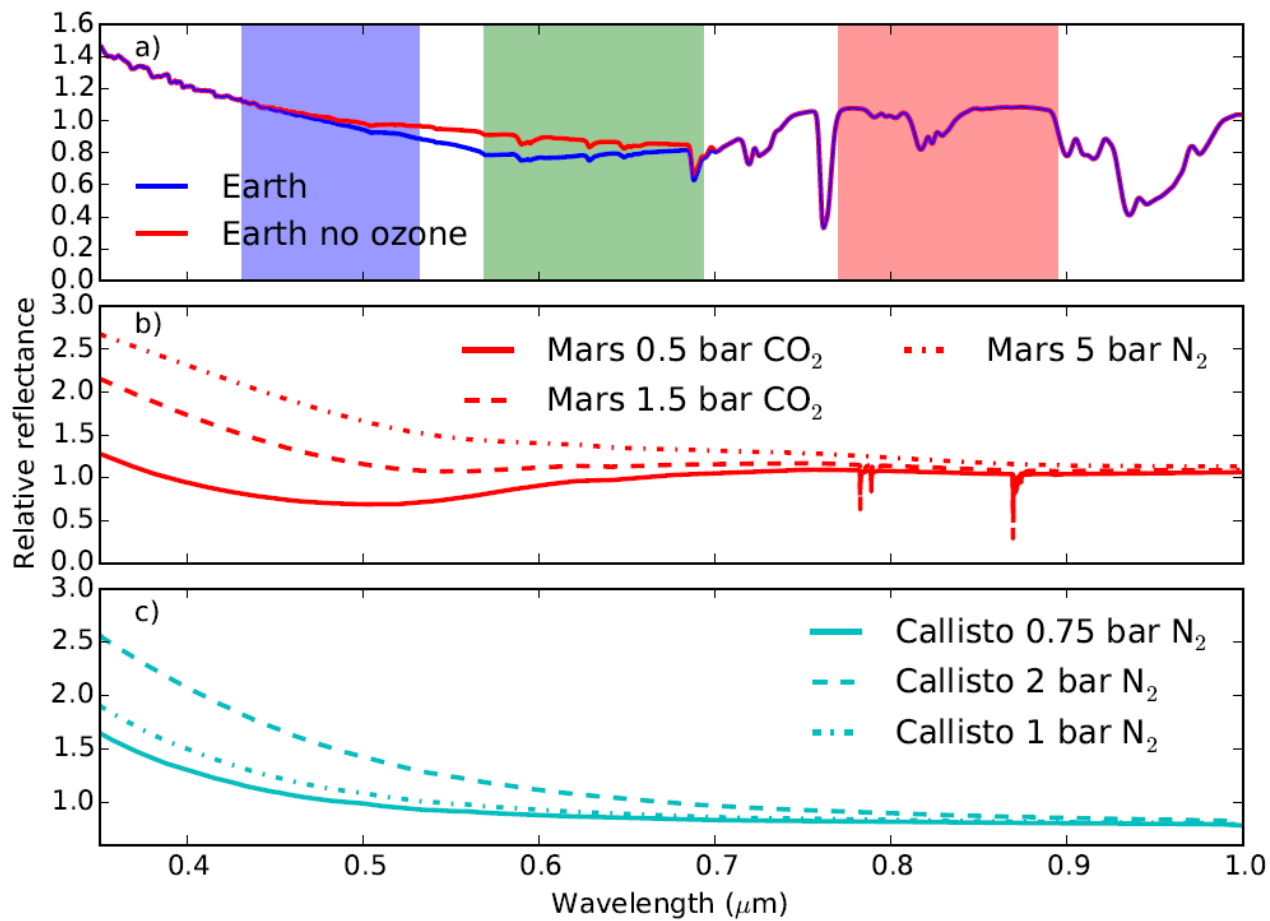

Figure 3: (a) shows Earth's reflectance spectrum (at quadrature) superimposed on the three optimal photometric bins from Fig. 2. These are the bins that maximize the separation between Earth and the nearest uninhabitable false-positive in color-color space. Note how the photometric bins highlight the U-shape of Earth's spectrum. For comparison we also plot Earth's spectrum without ozone to highlight how ozone absorption contributes to the unique U-shape. (b) Shows the reflectance spectra for a Mars-like surface with a thick atmosphere, and (c) shows reflectance spectra for Callisto with a thick $N_2$ atmosphere. These uninhabitable planets are mostly blue in color, but they don't have the same U-shape spectra as Earth, and so they are distinguishable using the optimal bins in (a).

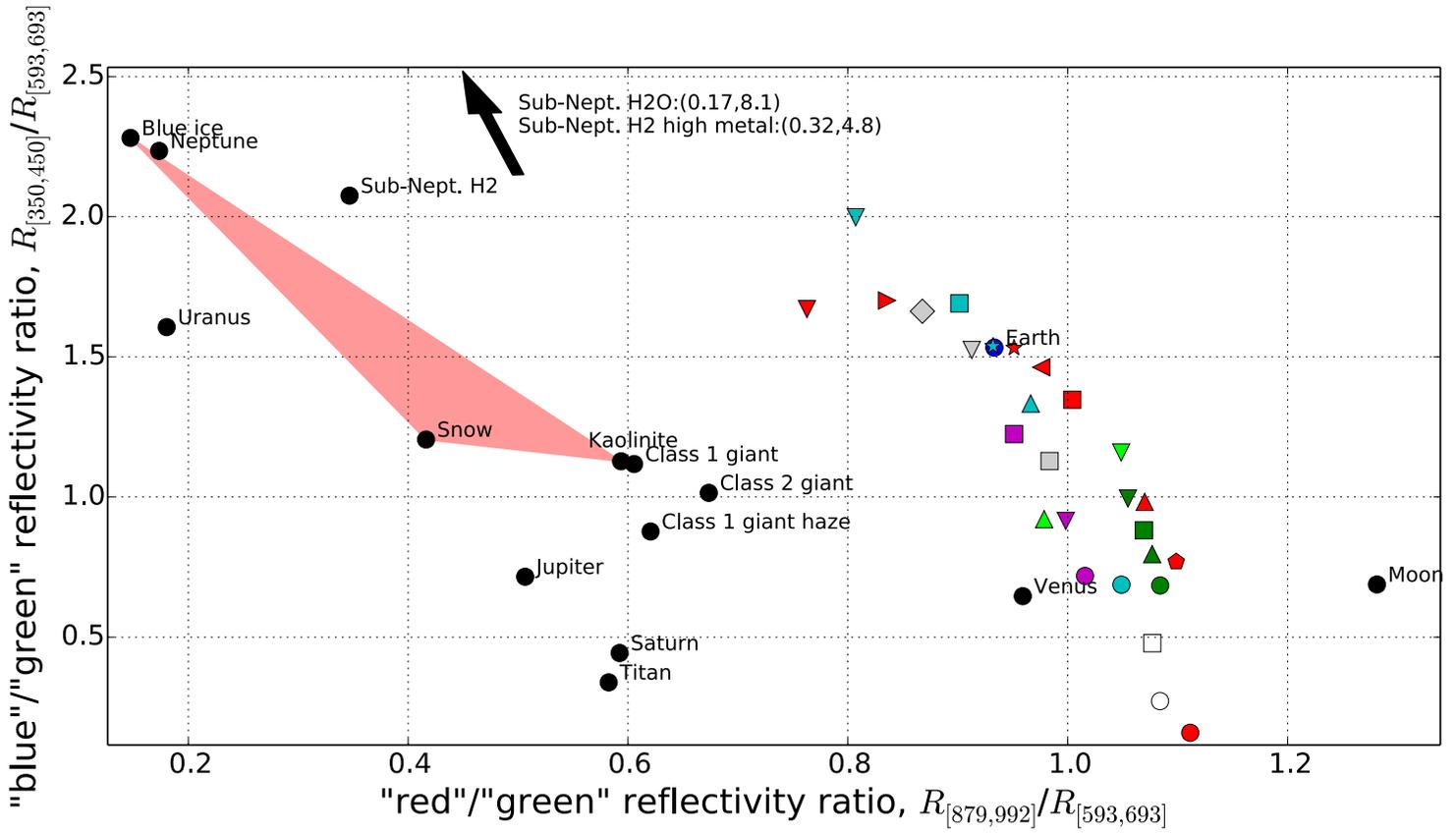

Figure 4: Here the Earth (blue circle) is optimally separated from all uninhabitable planets *except* for the icy false positives. Any planet with a circle symbol is included in the optimization calculation, whereas planets with other symbols (icy false positives) are not included. The legend is identical to that of Fig. 2. The horizontal axis is the mean reflectivity in the "red" bin (879-992 nm) divided by mean reflectivity in the "green" bin (593-693 nm), and the vertical axis is mean reflectivity in the "blue" bin (350-450 nm) divided by mean reflectivity in the "green" bin (593-693 nm). These photometric bins are optimized to maximize the separation between Earth and its nearest non-icy neighbor, a Class 1 giant (see text), and in this case the separation is 0.53. The arrow denotes two sub-Neptune models that fall beyond the axis limits of this figure, and the red shaded region represents the range of colors possible for a snowball Earth planet (see main text). Although Earth is better separated from non-icy false positives than in Fig. 2, it occupies the same region of color-color space as many icy bodies. In particular, Earth is identical in color to Mars with a 1.5 bar $CO_2$ atmosphere and Callisto with a 0.75 bar $N_2$ atmosphere using these photometric bins.

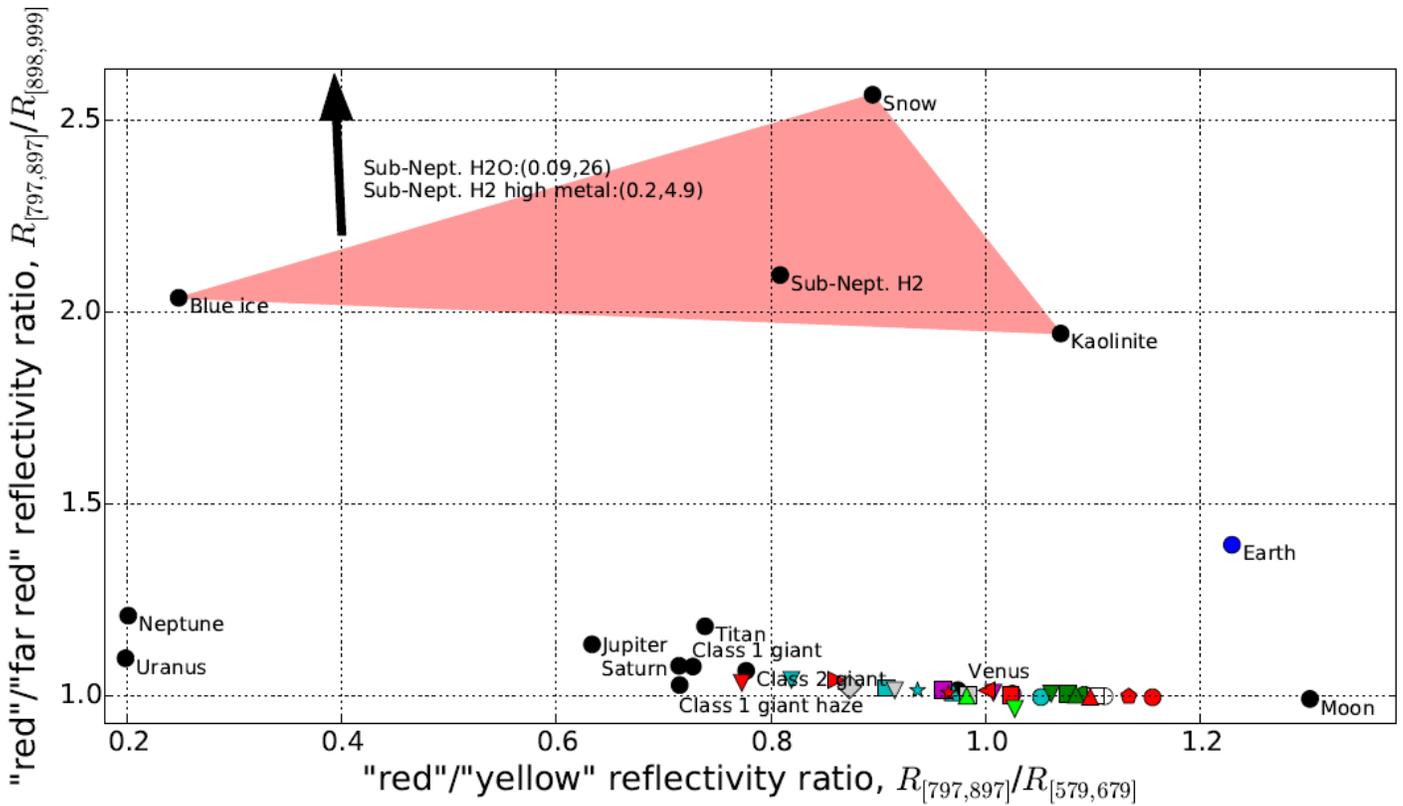

Figure 5: Here the Earth is optimally separated from all uninhabitable planets. Rather than enforce "red"/"green" and "blue"/"green" axes as in Fig. 2 and 4, in this figure the optimization algorithm considered all possible combinations of reflectivity ratios with three photometric bins. The three bins that define the axes are 579-679 nm ("yellow"), 797-897 nm ("red"), and 898-999 nm ("far red"), and the color-color space axes for this optimal separation are $R_{[797,897]} / R_{[579,679]}$ and $R_{[797,897]} / R_{[898,999]}$. The separation between Earth and its nearest neighbor (Mars) is 0.40. The arrow denotes two sub-Neptune models that fall beyond the axis limits of this figure, and the red shaded region represents the range of colors possible for a snowball Earth planet (see main text). The Earth has an unusually large ratio of $R_{[797,897]} / R_{[898,999]}$ due to water vapor absorption in its atmosphere (940 nm band).

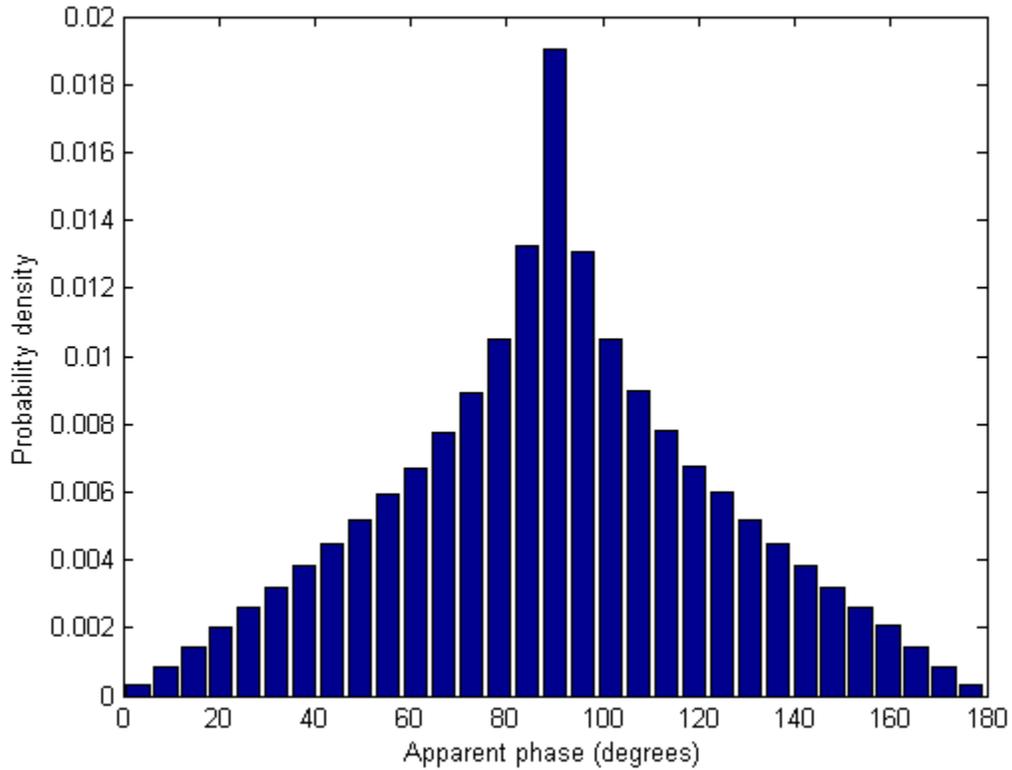

Figure 6: Probability distribution of the apparent phase for an exoplanet with a random position in its orbit and a random orbital inclination relative to the observer. This was calculated by repeatedly sampling uniform distributions of inclination and position in (a circular) orbit, and calculating the resultant phase as viewed by a distant observer. Quadrature (90 degrees) is the most likely apparent phase because planets with orbits inclined zero degrees relative to the Earth will always appear in quadrature. In practice, the distribution will be truncated at the edges due to inner working angle constraints.

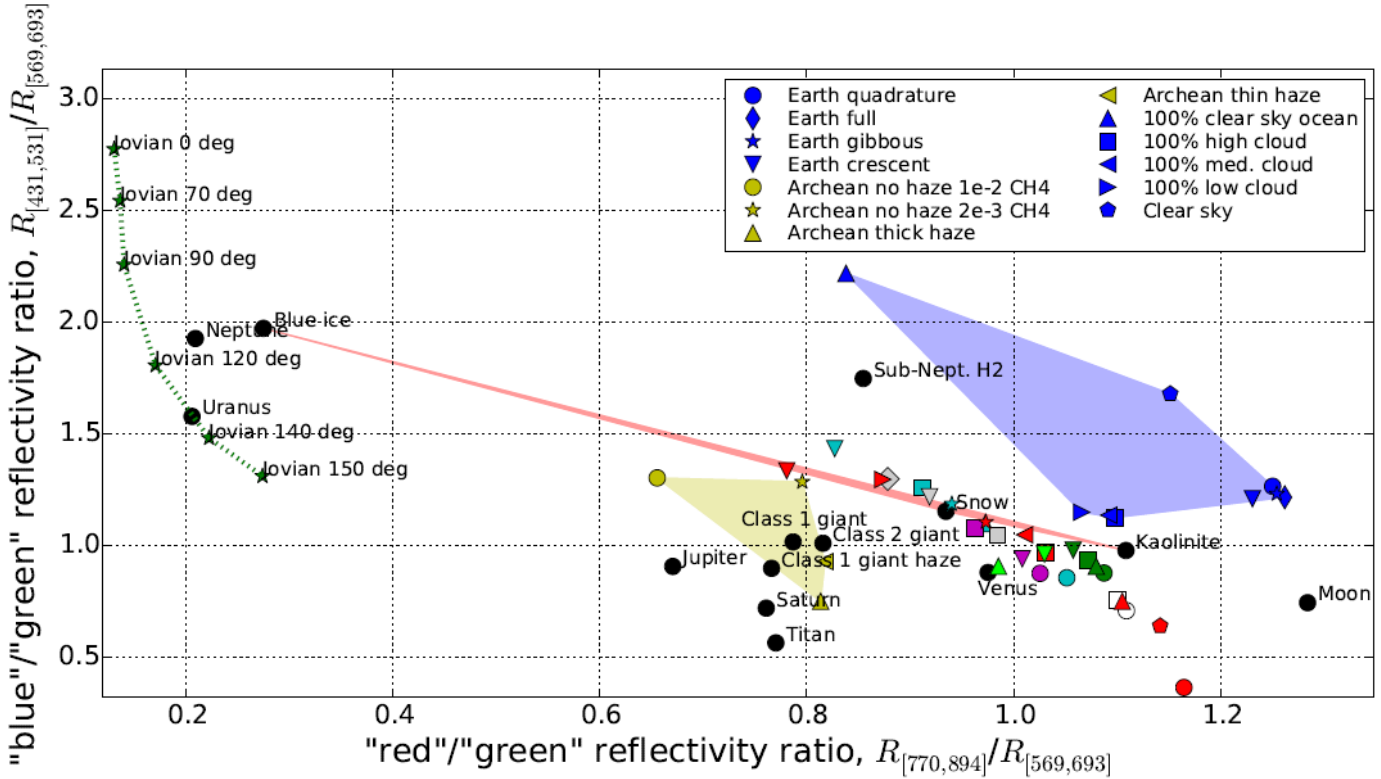

Figure 7: This figure is identical to Fig. 2 (including the legend) except a number of additional model planets have been added to illustrate the effects of phase, clouds, and haze. Note that these additional models are not optimally separated from Earth, but are instead merely plotted on top of the results from Fig. 2. The Earth in quadrature, gibbous, crescent and full phases (blue symbols) occupy very similar points in color-color space, indicating that phase does not have a large effect on the apparent color of Earth-like planets. However, phase can have a large effect on Jovian planets, as shown by the phase variation along the green dotted line. The reflectance spectra for this Jovian planet at 0.8 AU were obtained from Cahoy *et al.* (2010). To explore the effect of cloudiness on Earth's spectrum we have plotted Earth with 100% low, medium and high level cloud, in addition to a clear-sky Earth and a clear-sky Earth with a 100% ocean surface (blue symbols). The blue shaded region bounds these endmembers. Evidently, increased levels of cloudiness could make the Earth appear more similar in color to uninhabitable icy planets. Finally, we have plotted four Archean Earth spectra: thin haze, thick haze, no haze low methane, and no haze high methane (yellow symbols). The shaded yellow region represents the range of possible colors for the Archean Earth. Clearly, if these photometric bins were used to identify potentially habitable planets, then an exo-Archean Earth could be mistaken for a gas giant or Titan-like exoplanet.

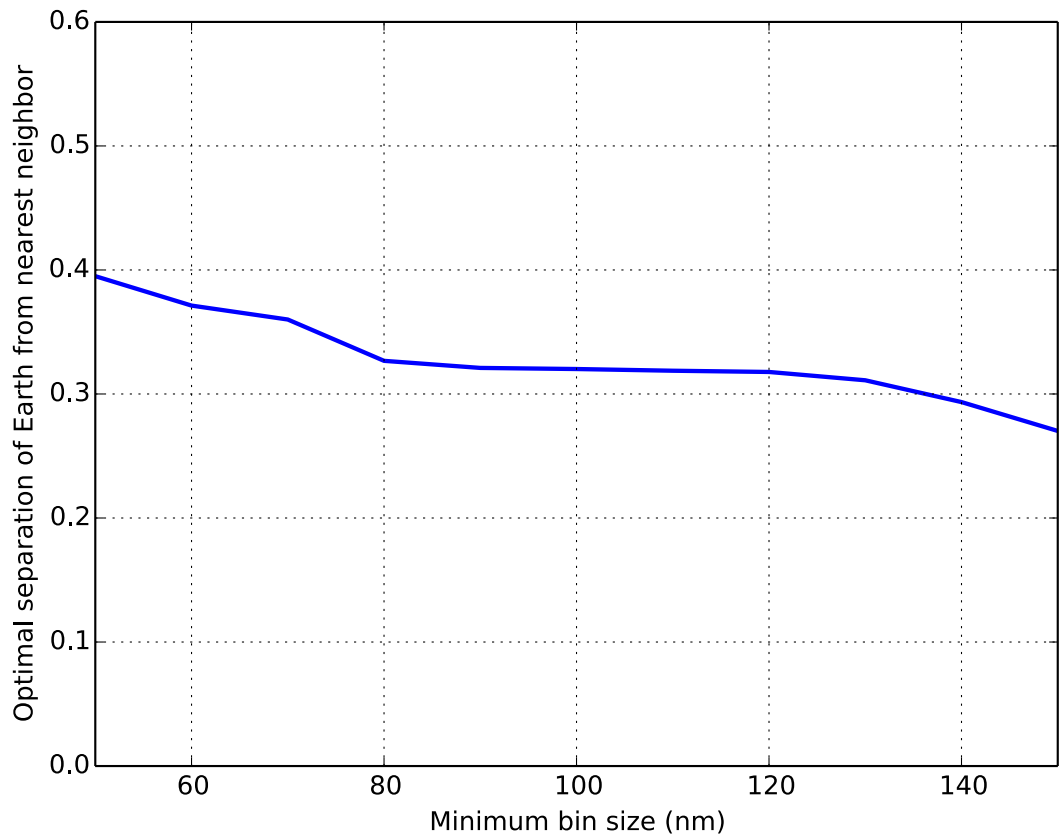

Figure 8: Optimal separation between Earth and its nearest uninhabitable neighbor (all false positives included) as a function of the minimum bin size constraint. Changing the minimum bin size does not have a dramatic effect on how well Earth can be separated.

9a

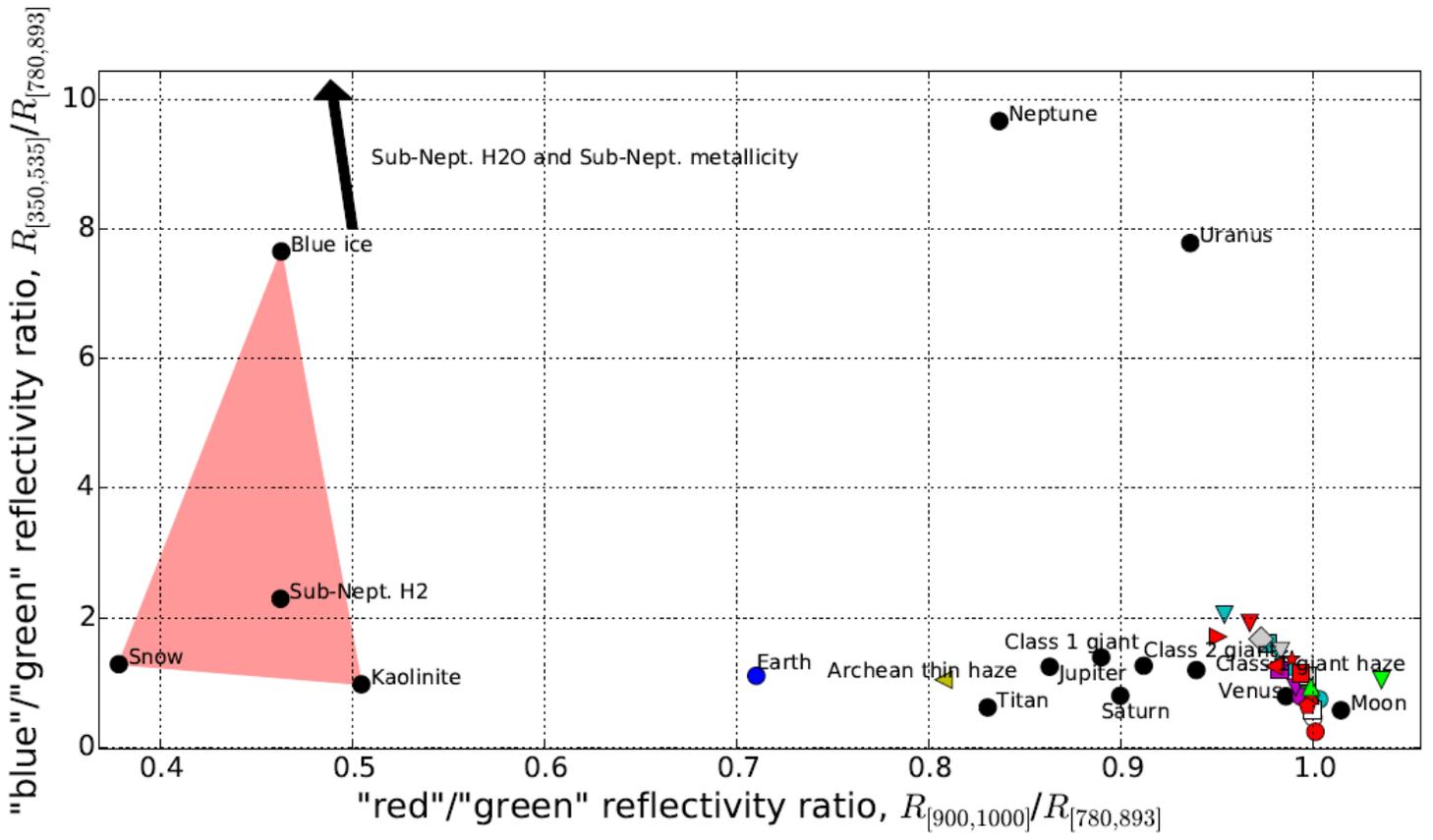

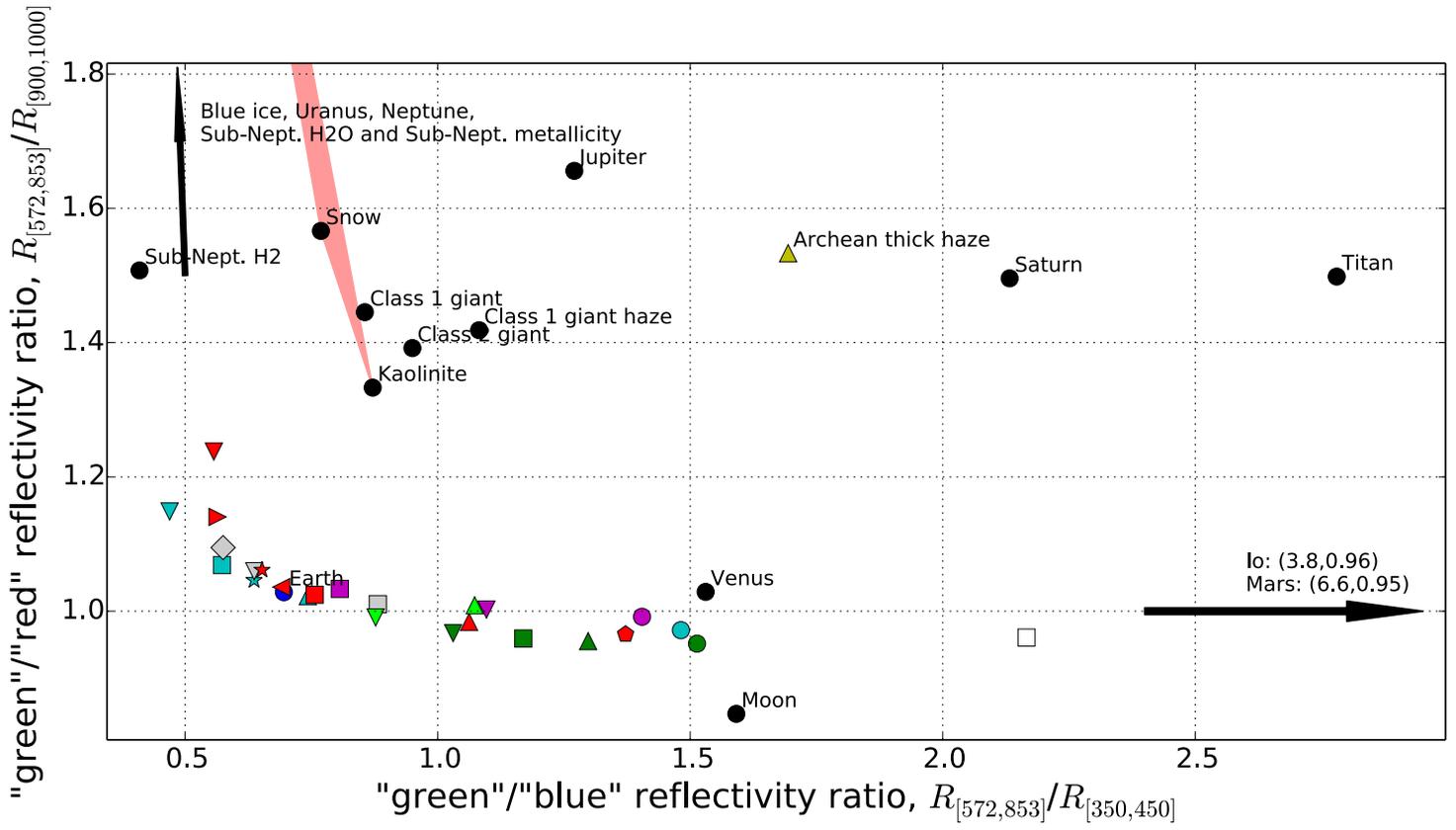

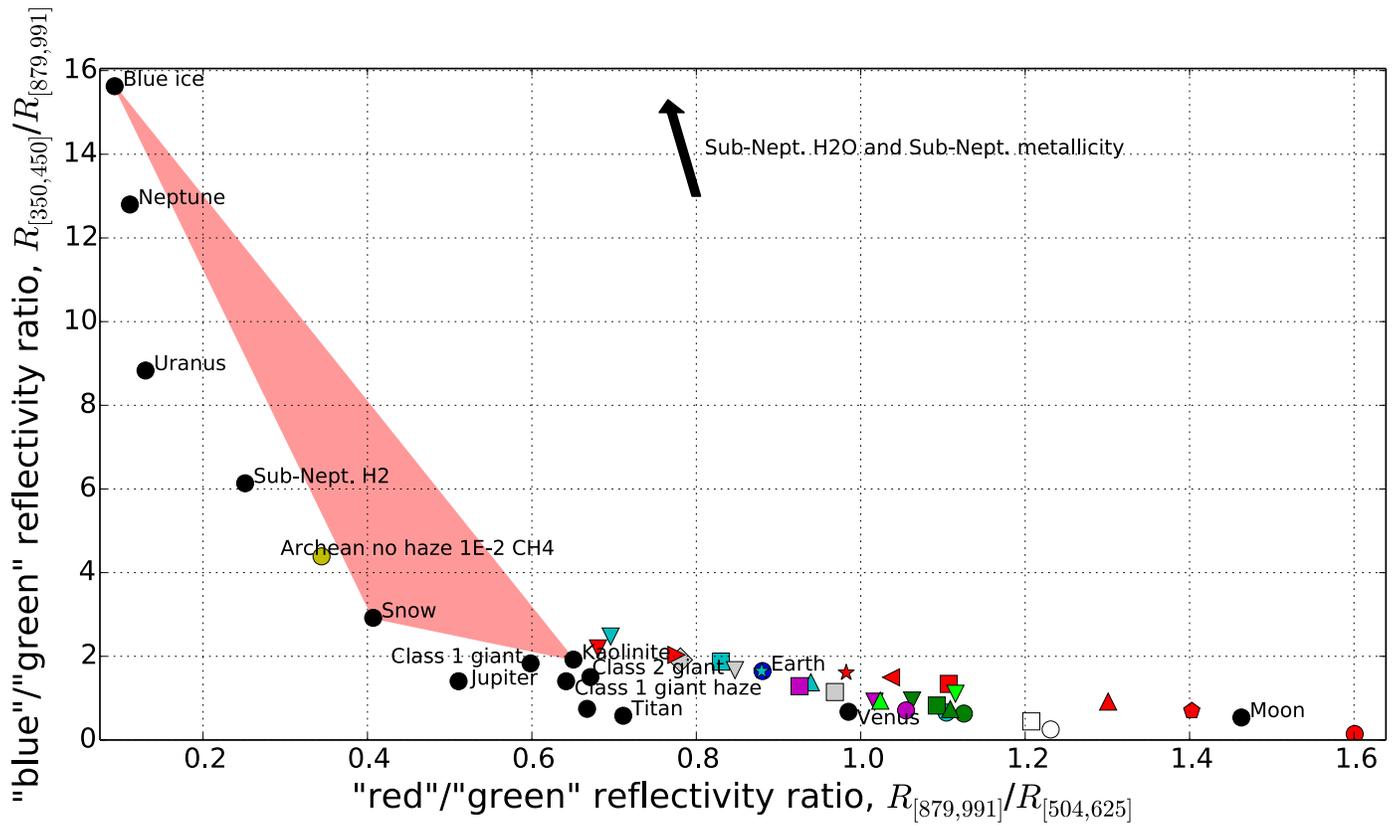

Figure 9: Here the Archean Earth has been optimally separated from all false positives. The legend is the same as Fig. 2, and the modern Earth is plotted for reference in each subfigure. (a) optimally separates the thin-haze Archean Earth from all false positives using "red"/"green" and "blue"/"green" reflectivity axes. The separation between the Archean Earth (left pointing yellow triangle) and its nearest neighbor, Jupiter, is only 0.20 (note that the scaling of the horizontal axis is bigger than the vertical axis). (b) optimally separates the thick-haze Archean Earth (upward pointing yellow triangle) from all false positives using generalized axes with three photometric bins. In this case the separation between the Archean Earth and its nearest neighbor is somewhat improved (0.44), but the Archean Earth does not occupy a unique region in color-color space since intermediates between Jupiter and Saturn are plausible. Finally, (c) optimally separates the non-hazy Archean Earth with high methane (yellow circle) from all false positives using generalized axes and three photometric bins. Once again the optimal separation is comparatively large (1.47), but the Archean Earth does not occupy a unique region in color-color space.

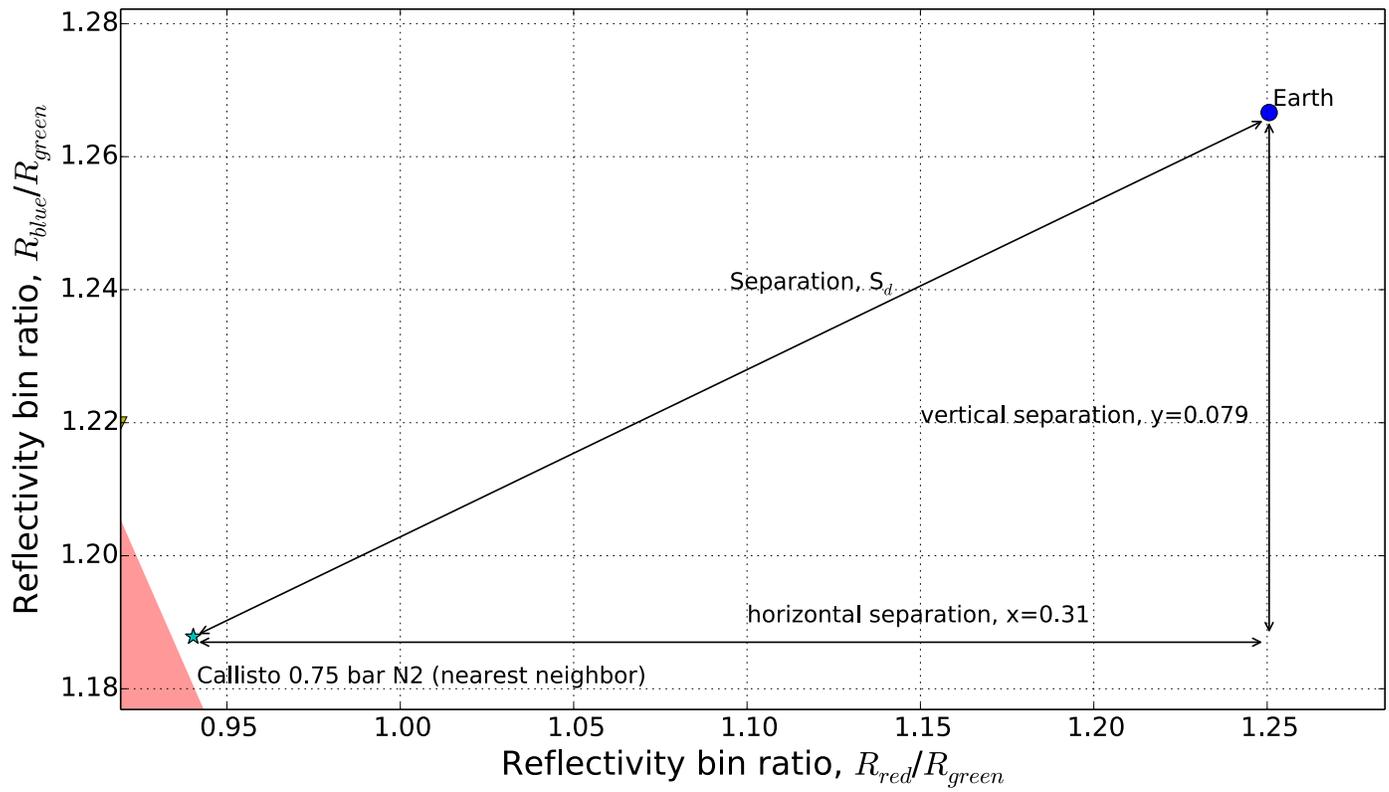

Figure 10: A zoomed in version of Fig. 2 illustrating the key parameters in the detectability calculations.